%% file: main_arxiv.tex
\documentclass[a4paper,12pt]{article}
\usepackage[paper=a4paper,left=25mm,right=25mm,top=25mm,bottom=25mm]{geometry} 
\pdfoutput=1

 \usepackage{setspace}
\usepackage{authblk}

\usepackage{fullpage}
\usepackage{parskip}
\usepackage{titlesec}
\usepackage[section]{placeins}
\usepackage[dvipsnames]{xcolor}

\usepackage{breakcites}
\usepackage{lineno}
\usepackage{hyphenat}
\usepackage[all]{nowidow}
\usepackage{longtable}

\usepackage{centernot} % für Befehl \centernot
\usepackage{lscape}

\PassOptionsToPackage{hyphens}{url}
\usepackage{hyperref} 
\hypersetup{colorlinks = true}

\hypersetup{
  linkcolor  = RedViolet,
  citecolor = Blue!90!black,
  urlcolor   = Aquamarine!80!black,
  colorlinks = true
}

\usepackage{etoolbox}
\usepackage{natbib}
\renewenvironment{abstract}
  {{\bfseries\noindent{\abstractname}\par\nobreak}\footnotesize}
  {\bigskip}
\titlespacing{\section}{0pt}{*3}{*1}
\titlespacing{\subsection}{0pt}{*2}{*0.5}
\titlespacing{\subsubsection}{0pt}{*1.5}{0pt}

\usepackage{graphicx}
\usepackage[space]{grffile}
\usepackage{latexsym}
\usepackage{textcomp}
\usepackage{tabulary}
\usepackage{booktabs,array,multirow}
\usepackage{amsfonts,amsmath,amssymb}
\usepackage{subcaption}
\usepackage{booktabs} % für toprule, midrule und bottomrule 
\usepackage{tabularx}
\usepackage{amssymb}
\usepackage{tabularx}
\usepackage{pifont}
\usepackage{caption} % to center a caption
\usepackage{graphicx}
\usepackage{hyperref}
\usepackage{comment}
\providecommand\citet{\cite}
\providecommand\citep{\cite}

\newif\iflatexml\latexmlfalse

\AtBeginDocument{\DeclareGraphicsExtensions{.pdf,.PDF,.eps,.EPS,.png,.PNG,.tif,.TIF,.jpg,.JPG,.jpeg,.JPEG}}

\usepackage[utf8]{inputenc}
\usepackage[english]{babel}

\usepackage{orcidlink}
\newcommand{\beginsupplement}{
        \setcounter{section}{0} 
        \renewcommand{\thesection}{S\arabic{section}}

        \setcounter{table}{0}
        \renewcommand{\thetable}{S\arabic{table}}
        \setcounter{figure}{0}
        \renewcommand{\thefigure}{S\arabic{figure}}
     }

\newcommand{\tool}[1]{\texttt{#1}}

\newcommand{\newcrossmark}{\scalebox{1}[1]{$\times$}}

\definecolor{bleudefrance}{rgb}{0.19, 0.55, 0.91}

\newcommand{\tabitem}{~~\llap{\scalebox{0.7}\textbullet}~~}

\usepackage{array}
\newcolumntype{Y}[1]{>{\centering\arraybackslash}p{#1}}

\begin{document}
\title{From RNA sequencing measurements to the final results: a practical guide to navigating the choices and uncertainties of gene set analysis}

\def\correspondingauthor{\footnote{Corresponding author, e-mail: \href{mailto:milena.wuensch@ibe.med.uni-muenchen.de}{milena.wuensch@ibe.med.uni-muenchen.de}, Institute for Medical Information Processing, Biometry, and Epidemiology, LMU Munich, Marchioninistr. 15, D-81377, Munich, Germany.}}
\author[1,2]{Milena Wünsch \correspondingauthor{} \orcidlink{0009-0001-1982-9260}}
\author[1,2]{Christina Sauer \orcidlink{0000-0003-2425-7858}}
\author[1]{Patrick Callahan \orcidlink{0000-0003-1769-7580}}
\author[3]{Ludwig Christian Hinske \orcidlink{0000-0001-7273-5899}}
\author[1,2]{Anne-Laure Boulesteix \orcidlink{0000-0002-2729-0947}}
\affil[1]{Institute for Medical Information Processing, Biometry, and Epidemiology, LMU Munich, Munich (Germany)}
\affil[2]{Munich Center for Machine Learning, Munich (Germany)}
\affil[3]{Institute for Digital Medicine, University Hospital of Augsburg, Augsburg (Germany)}
\vspace{-1em}
  \date{\today}
\maketitle

%\author[1,2,$\ast$]{Milena Wünsch \ORCID{0009-0001-1982-9260}}
%\author[1,2]{Christina Sauer \ORCID{0000-0003-2425-7858}}
%\author[1]{Patrick Callahan \ORCID{0000-0003-1769-7580}}
%\author[3]{Ludwig Christian Hinske}
%\ORCID{0000-0001-7273-5899}
%\author[1,2]{Anne-Laure Boulesteix}
%\ORCID{0000-0002-2729-0947}

\begin{abstract}

Gene set analysis, a popular approach for analyzing high-throughput gene expression data, aims to identify sets of related genes that show significantly enriched or depleted expression patterns between different conditions. In the last years, a multitude of methods and corresponding tools have been developed for this task. However, clear guidance is lacking: choosing the right method is the first hurdle a researcher is confronted with. No less challenging than overcoming this so-called method uncertainty is the procedure of preprocessing, from knowing which steps are required to selecting a corresponding approach from the plethora of valid options to create the accepted input object (data preprocessing uncertainty), with clear guidance again being scarce. Here, we provide a practical guide through all steps required to conduct gene set analysis, beginning with a concise overview of a selection of established methods, including GSEA and DAVID. We thereby lay a special focus on reviewing and explaining the necessary preprocessing steps for each method under consideration (e.g. the necessity of a transformation of the RNA-Seq data)---an essential aspect that is typically paid only limited attention to in both existing reviews and applications. To raise awareness of the spectrum of uncertainties, our review is accompanied by an extensive overview of the literature on valid approaches for each step and illustrative R code demonstrating the complex analysis pipelines. It ends with a discussion and recommendations to both users and developers to ensure that the results of gene set analysis are, despite the above-mentioned uncertainties, replicable and transparent.

\end{abstract}

\sloppy

\input{01_introduction}
\input{02_general_principle}
\input{03_methods_tools}

\input{04_preprocessing}
\input{05_implications}

\section*{Illustrations in R}
Alongside this review, we provide an illustration of the practical steps to be performed in \texttt{R} for each step from Figure~\ref{fig_pre_workflow}. For an overview of the structure of the illustration, refer to Section 6 in the supplement. For this illustration, the user has the choice to work through several \texttt{R} files (available on \href{https://github.com/chillemille/RIllustrations}{\color{bleudefrance}{GitHub}}), however, they can also inspect them visually as a part of \href{https://chillemille.github.io/RIllustrations/}{\color{bleudefrance}{A Practical Guide through Running Gene Set Analysis in R}}.

\section*{Funding Information}

{\label{974317}}
This work is supported in part by funds from the German Research Foundation (DFG: BO3139/7-1 and BO3139/9-1). 

\section*{Acknowledgements}

{\label{749861}}

The authors thank Savanna Ratky for language correction.

\section*{Competing interests}
No competing interest is declared.

\FloatBarrier
\bibliographystyle{apalike}
\bibliography{reference}

\newpage
\beginsupplement
\input{06_supplementary_material}

%\FloatBarrier
%\bibliographystyle{apalike}
%\bibliography{reference}
\end{document}

%% file: 01_introduction.tex
\section{Introduction}

Gene set analysis (GSA) is a common approach to gaining insight into high-throughput gene expression data by detecting sets of related genes that show an enriched or depleted expression pattern between two conditions (such as cases and controls of a specific disease). GSA is thus considered an extension to the no less well-known differential expression analysis \citep{khatri2012ten}, which, on the other hand, produces a list of individual genes that show a significant difference in gene expression between two conditions. Through the aggregation of genes with a common relation into gene sets, GSA boasts an increased statistical power and a simplified interpretation of the results, compared to differential expression analysis \citep{ackermann2009general, khatri2012ten}. 

While a multitude of different methods and corresponding tools have been proposed for GSA in the last years, it is hard for benchmark studies to keep pace with these developments. Moreover, it has been observed that these benchmark studies more often rely on subjective assessments of the consistency of the methods' results with existing biological knowledge than on well-designed simulation studies \citep{xie2021popularity}. This lack of reliable guidance for users aiming to choose the most suitable GSA method results in what we denote as {\it method uncertainty}, following the terminology introduced by \citet{hoffmann2021multiplicity} in a more general context. 

This method uncertainty is accompanied by a variety of options within each individual GSA method, such as the choice of particular numeric parameters or the choice of the gene set database. These options in principle enable the methods' users to adapt them to the given research question and data at hand. While loose guidance is provided in the form of a default option for some parameters, the choice of the most suitable parameter value is often unclear, resulting in what we denote as {\it parameter uncertainty}. 

The lack of guidance is even more pronounced when it comes to the generation of the required input objects of the individual GSA methods (e.g. list of differentially expressed genes, ranked list of all genes, or preprocessed gene expression data set). Here, the choice of the exact approach to preprocessing among a variety of valid options is largely left to the user, typically without any guidance through default approaches. Particularly for users with little bioinformatics experience, this so-called {\it data preprocessing uncertainty} may present a considerable difficulty when conducting GSA. 

While a variety of review articles addressing GSA have been published in the past, they primarily focus on the theoretical principles underlying the methods. For instance, some summarize the underlying structures of existing GSA methods into a common modular framework \citep{ackermann2009general, maleki2020gene}. Others discuss the challenges and limitations of the three general approaches to GSA \citep{khatri2012ten}. Practical guidance for GSA is provided for two web-based tools \citep{reimand2019pathway}, however, focusing primarily on using the tools and interpreting the results while touching only sparsely upon other choices related to data preprocessing. 
 
In the general context of statistical analyses, the entirety of choices the user is confronted with is commonly referred to as researchers' degrees of freedom \citep{Simmons2011}. In the context of GSA, the researchers' degrees of freedom related to method, parameter, and data preprocessing uncertainty outlined above may make it difficult for users to obtain an overview of all steps required to conduct the analysis. This circumstance is further aggravated by the varying quality of the instructions and user manuals provided for the different methods. While for some methods, a detailed illustration of the required steps for preprocessing and running the method itself is provided, the manuals of other methods are considerably less instructive. 

In view of this situation, the aim of this paper is to bring light into the darkness of all steps required to conduct GSA. While we cannot solve the spectrum of choices and uncertainties the user is confronted with by giving explicit recommendations among the variety of options, we believe that the awareness of their existence and relevant characteristics is an important step towards more reliable results. The aim of our paper is three-fold. Firstly, we provide an overview of a selection of methods and tools identified as widely-used and/or well-performing in previous literature \citep{xie2021popularity} including criteria guiding the choice (Sections `General Principle' and `Methods and Tools'). Secondly and most importantly, we review the practical steps related to preprocessing procedures required to generate the input objects accepted by the considered GSA tools (Section `Preprocessing and Practical Aspects'). This part is illustrated through documented example \texttt{R} codes. Thirdly, we discuss the implications of the different types of uncertainties and formulate recommendations for methods' users and developers (Section `Implications and Recommendations'). In the supplement, we address existing parameter and data preprocessing uncertainties by providing an extensive review of the multitude of available options and approaches. 

%% file: 02_general_principle.tex
\section{General Principle}\label{sec2}
Methods developed for GSA typically have the same starting point and follow (variants of) a common framework. An understanding of this framework forms the cornerstone for recognizing similarities and differences between the methods and choosing the most suitable one for a specific research question at hand. 
In the following, we give an overview of the starting point of all GSA methods which is followed by a description of two general approaches to GSA. 

%%%%%%%%%%%%%%%%%%%%%%%%%%%%%%
%%% Figure about GSA approaches
%%%%%%%%%%%%%%%%%%%%%%%%%%%%%%

\begin{figure*}[ht]
\centering
       
\includegraphics[scale=0.8]{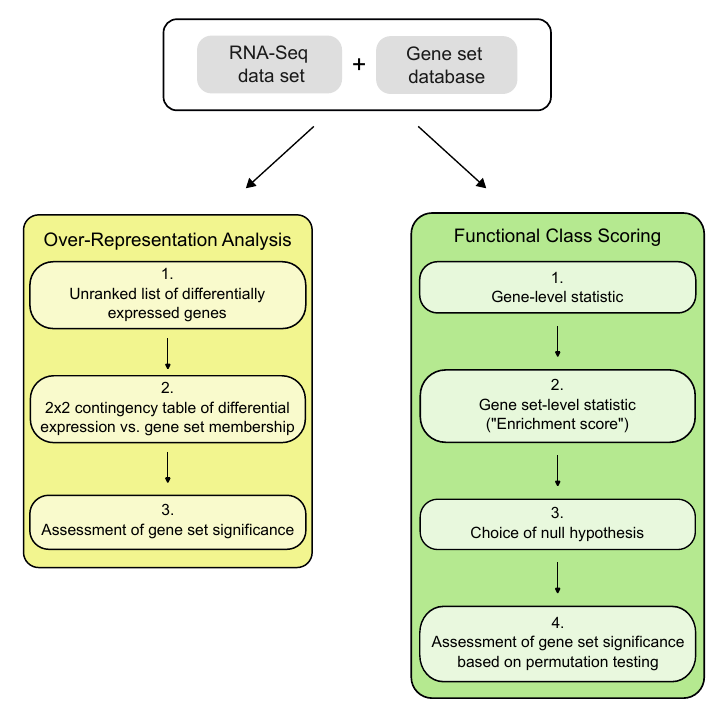}
\caption{\small General overview of GSA approaches ORA and FCS, their required input, and preparatory steps. Depending on the form of the expression data set and the respective GSA tool, additional preparatory steps might be necessary. The colors are in accordance with Figure~\ref{fig_pre_workflow}.}\label{fig_gsa_overview}
        \end{figure*}

Two components are required for GSA, namely i) a gene expression data set, with corresponding assignments of the samples to the conditions, such as case and control of a specific disease, and ii) a gene set database. The gene expression data set of dimension $N \times p$ contains the individual gene expression measurements of $N$ genes across $p$ samples in the form of integer count data. In principle, a higher value indicates a higher level of gene expression (see Section `RNA-Seq data set' in the supplement for a more detailed description). The second component that is required for GSA is a gene set database which provides information on how the genes from the experiment are aggregated into gene sets. This aggregation is typically based on specific commonalities between the genes such as common chromosomal locations or biological functions \citep{subramanian2005gene}. We note that individual genes can, dependent on the gene set database, be assigned to more than one gene set, resulting in a certain overlap between some gene sets.\\
Statistical methods for GSA can be divided into three major approaches, namely Over-Representation-Analysis (ORA), Functional Class Scoring (FCS), and Pathway Topology (PT). These approaches differ in the extent to which the information from the gene expression data set is utilized and therefore greatly vary in the complexity of the underlying methodology.\\
In this work, we focus on methods and tools ascribed either to ORA or FCS for which an overview of the underlying methodology is provided in Figure~\ref{fig_gsa_overview}. This focus is substantiated by the selection criterion described in Section `Methods and Tools'. However, for completeness, a short description of methods assigned to PT is provided in Section 1.2 in the supplement.

%%%%%%%%%%%%%%%%%%%%%%%%%%%%%%%%%%%%%%%%%%%%%%%%%%%%%%%%%%%%%%%%%%%%%%%%%%%%%%%%%%%%%%%%%
\subsection{Over-Representation Analysis (ORA)}
GSA methods classified as ORA \citep{draghici2003global}, which can be found on the left part of Figure~\ref{fig_gsa_overview}, commonly are the least complex among the three approaches. The required input typically consists of a list of differentially expressed genes which is generated using a suitable method for differential expression analysis (step~1). Then, a contingency table is generated which displays the frequencies of differential expression (differentially expressed versus not differentially expressed) and membership to the given gene set (member of the given gene set versus not a member of the given gene set) for all genes from the experiment (step~2). For an illustration of the corresponding contingency table see Table \ref{tab:cont_table_ora}. \\
The null hypothesis states that the given gene set is not differentially enriched, i.e. that there is no association between membership of a given gene set and differential expression. Accordingly, a specific gene set is detected as differentially enriched if the number of its gene set members that are differentially expressed ($H$) is too high to be caused solely by chance (step~3). The null distribution of $H$ is modeled using the hypergeometric distribution. In the context of ORA, the corresponding population size (i.e. the entirety of genes) is typically referred to as `universe' or `set of background genes'. Note that an elaboration on important aspects of the universe is provided in Section 4.1.2 in the supplement. Eventually, a $p$-value of over-representation, on which the assessment of differential enrichment of the given gene set is based, is typically obtained using Fisher's exact test, which is described in further detail in Section 1.1 in the supplement. \\

\begin{table*}
\centering
\caption{\small Contingency table in Over-Representation Analysis generated from all genes in the experiment. N: Total number of genes from the experiment; H: Number of genes from input that are members of the given gene set (`hits'); L: number of genes in the input list, G: number of genes that are members of the given gene set.}
\label{contingencytable_ora}
\begin{tabularx}{0.9\textwidth}{p{0.2\textwidth} |Y{0.25\textwidth} Y{0.25\textwidth}|p{0.2\textwidth}}
 & Differentially expressed & Not differentially expressed & Total \\
\hline
Member of \newline gene set & $H$ & $G - H$& $G$ \\
Not a member \newline  of gene set & $L-H$ & $N-L-(G-H)$ &$N-G$ \\
\hline
Total & $L$ & $N-L$& $N$\\
\end{tabularx}
\label{tab:cont_table_ora}
\end{table*}

%%%%%%%%%%%%%%%%%%%%%%%%%%%%%%%%%%%%%%%%%%%%%%%%%%%%%%%%%%%%%%%%%%%%%%%%%%%%%%%%%%%%%%
\subsection{Functional Class Scoring (FCS)}
Unlike ORA, methods and tools classified as FCS (see right part of Figure~\ref{fig_gsa_overview}) do not only consider the subset of differentially expressed genes but instead utilize the expression profiles of the entirety of genes from the experiment \citep{maleki2020gene}. For each gene, a gene-level statistic is generated which reflects the extent to which its expression pattern differs between both conditions (step~1) \citep{ackermann2009general}. Here, a positive and negative value indicate an association with the first and second condition, respectively. A corresponding ranking of all genes from the experiment is generated based on their values of the gene-level statistic. In this ranking, a position at the upper and lower tail of the ranking indicates a strong association of the corresponding gene with the first and second condition, respectively, while a location in the middle implies that there is no particular association with any of the conditions. While some FCS methods generate this gene ranking internally, meaning that they only require the input of a (preprocessed) gene expression data set, others require the user to provide a gene ranking that has been generated externally.\\
Then, for a given gene set, the values of the gene-level statistics (or in some cases, the corresponding ranks) of all gene set members are aggregated into a gene set-level statistic (typically called enrichment score) to summarize whether the gene set members' expression patterns gravitate towards one of both conditions, i.e. the genes are located towards the upper or lower end of the ranking, or else not associated with either of the conditions (step~2) \citep{subramanian2005gene}.\\ 
There are two types of null hypotheses that determine how to assess differential enrichment of a given gene set (step~3), namely competitive and self-contained \citep{goeman2007analyzing}. Depending on the type of the null hypothesis, a $p$-value is obtained through an appropriate procedure of permutation that empirically generates a null distribution of enrichment scores against which the true enrichment score is compared (step~4). The choice of the competitive null hypothesis implies a comparison of the association between the given gene set and the conditions to the association between the remaining genes and the conditions. In this case, the null distribution of enrichment scores is in practice approximated by repeating the analysis a large number of times with randomly generated fictive gene sets from the entirety of genes in the experiment that are of the same size as the given one, a procedure denoted as gene set permutation. For each of these randomly generated gene sets, an enrichment score is obtained from the initial ranking of all genes. \\
In contrast, a self-contained null hypothesis focuses on the given gene set, regardless of the remaining genes in the experiment. Accordingly, the null distribution of enrichment scores is obtained by generating a large number of random permutations of the assignments of the conditions to the samples (phenotype permutation). For each of these permutations, a corresponding gene ranking is generated from which the enrichment score is extracted. \\

%% file: 03_methods_tools.tex
\section{Methods and Tools}
Over the past years, a multitude of methods and corresponding tools have been proposed for both ORA and FCS which differ in the specifics to assess differential enrichment. Beyond the distinction of ORA versus FCS, these may include biological aspects considered to be relevant as well as the handling of characteristics inherent to RNA-Seq data which might affect the results. \\
Note that in this paper, we explicitly distinguish between GSA \textit{methods} and \textit{tools} under the aspect that each \textit{method} is implemented in one or more \textit{tools}, either exactly or with some modifications. For instance, the common method Gene Set Enrichment Analysis \citep{subramanian2005gene} has been implemented in several tools, namely directly in a web-based tool but also, with slight variations to the original underlying methodology, in several \texttt{R} packages. \\
From the multitude of existing methods and tools, we have made a selection based on popularity that is presented in Table \ref{overview_methods_tools}. 
This selection is derived from the work of \citet{xie2021popularity} who provide a comprehensive reference database of existing GSA methods and tools together with a quantification of the respective popularity. For a more detailed description of the selection criterion, inspect Section 2.1 in the supplement. Note that the resulting selection consists only of methods and tools assigned to ORA or FCS since PT generally scores considerably lower in terms of popularity in the considered reference database. \\
In the following, we provide a compact summary of each method and tool among this selection and focus on their corresponding specifics. For each of these tools, we additionally illustrate the correct use in the form of \texttt{R} code. Additionally, we address the parameter uncertainties by demonstrating in the illustrations how the corresponding parameters can be adapted in the code.

\begin{table*}[ht]
\small
\centering
    \caption{\small Overview of the considered GSA methods and tools, considering that each method (`Method') is implemented in one or more associated tools (`Implemented in tool(s)'). While the column `\texttt{R} or Web' indicates whether the associated GSA tool is implemented in \texttt{R} or a web-based tool, the column `Selection criterion' distinguishes between a high popularity or a high performance as the criterion used to select the tool. Note that, while we use the written out term to refer to the method `Gene Set Enrichment Analysis', we use the corresponding abbreviation `\tool{GSEA}' for its implementations (i.e. tools) to avoid confusion.}
    \begin{tabularx}{0.95\textwidth} {p{0.25\textwidth} p{0.1\textwidth}p{0.19\textwidth}p{0.15\textwidth}p{0.15\textwidth}}
    \toprule
      Method & ORA/ \newline FCS  & Implemented \newline in tool(s) & \texttt{R} or Web & Selection criterion  \\
         \midrule
        ORA &  ORA & \tool{DAVID} \newline \texttt{clusterProfiler} & Web \newline \texttt{R} & Popularity \newline Popularity \\ 
        GOSeq & ORA & \tool{GOSeq}  & \texttt{R} & Popularity\\
         Gene Set Enrichment Analysis & FCS & \tool{clusterProfiler} \newline \tool{GSEA} \newline \tool{GSEAPreranked} & \texttt{R} \newline Web  \newline  Web  & Popularity \newline Popularity \newline Popularity \\
         PADOG  & FCS &  \tool{PADOG} & \texttt{R} & Performance\\
        
        \bottomrule
        
    \end{tabularx} \\ 

    \label{overview_methods_tools}
\end{table*}

\subsection{Methods and Tools for Over-Representation Analysis}
In the following, we present several popular methods and tools developed for ORA. We further elaborate on existing parameter uncertainties, i.e. common flexible parameter choices within the corresponding ORA tools, in Section 4.1 in the supplement.

\subsubsection{\texttt{clusterProfiler}'s ORA}
In contrast to DAVID and GOSeq, which apply some modifications to the general ORA framework (see Section `Over-Representation Analysis (ORA)'), a direct implementation is offered by the \texttt{R} package \tool{clusterProfiler} \citep{wu2021clusterprofiler}. Given the list of differentially expressed genes as input, a two-dimensional contingency table (see Table~\ref{tab:cont_table_ora}) is generated which displays the frequency of differentially and non-differentially expressed genes in one dimension and the frequency of genes that are members of the given gene sets and genes that are not in the second dimension. Eventually, the $p$-value of over-representation of a given gene set is assessed using the hypergeometric distribution and Fisher's exact test.

\subsubsection{DAVID}\label{david}
Database for Annotation, Visualization, and Integrated Discovery (\tool{DAVID}) \citep{huang2009bioinformatics, huang2009systematic} is a collection of web tools developed to provide an understanding of the biological meaning behind lists of genes. Within this collection of tools, the functional annotation tool provides a conservative variant of the classical ORA framework presented above such that the (adjusted) $p$-value increases for each gene set. While gene sets with fewer members are affected more strongly by this increase compared to larger gene sets, a significant over-representation of gene sets containing only a single gene set is precluded completely.

\subsubsection{GOSeq}
GOSeq \citep{young2010gene} is an ORA methodology that addresses the circumstance that for RNA-Seq data, the probability of detecting a gene set as differentially enriched increases as the gene set members increase in transcript length, even if they remain constant in their level of differential expression between both conditions \citep{oshlack2009transcript}. A more detailed description of this so-called `length bias' is provided in Section 2.3 in the supplement. \\
GOSeq counteracts length bias by incorporating each gene's estimated probability of being detected as differentially expressed depending on the gene's transcript length in the form of a probability weighting function. \\
By default, Wallenius' non-central hypergeometric distribution \citep{wallenius1963biased}, which employs an extension of the standard hypergeometric distribution, is used to assess differential enrichment. In this method is assumed that within the given gene set, all genes have the same probability of being drawn from the universe and that this probability differs from the common probability of all genes outside of this gene set.

\subsection{Methods and Tools for Functional Class Scoring}
Before giving a short overview of each FCS method and tool under consideration, we will elaborate on an important distinction between FCS tools which has a relevant impact on the practical steps that have to be performed by the user. Additionally, we elaborate on common flexible parameter choices for these tools (i.e. parameter uncertainties) in Section 4.2 in the supplement.

\subsubsection{Distinction: FCS~I vs. FCS~II}
One substantial difference between different tools classified as FCS is the form of the input object. While some tools require a manually ranked gene list as input, others accept a (preprocessed) gene expression data set as a whole, from which the gene ranking is generated internally. Since this distinction leads to a noticeable difference in the required preprocessing, it is addressed in this work by the subdivision of the FCS tools (see Table \ref{fcs_distinction}) into FCS~I, which contains all tools that accept as input the (preprocessed) gene expression data set, and FCS~II, which is based on a gene ranking. Note that for FCS~II, this gene ranking is typically generated by conducting differential expression analysis and applying a suitable gene-level statistic to the quantities in the corresponding results table (see Section `Gene-Level Statistic'). \\
Since in the transition from the gene expression data set to the gene ranking, the information of the conditions (i.e. phenotypes) of the samples is lost, phenotype permutation, which is accompanied by a self-contained null hypothesis, cannot be performed for FCS~II methods. Therefore, only gene set permutation (and thus a competitive null hypothesis) can be used to assess the significance of a given gene set. \\
We underline that the distinction between FCS~I and FCS~II is only made between FCS tools (and not FCS methods). For instance, the {\it tool} \tool{GSEA} is classified as FCS~I and \tool{GSEAPreranked} as FCS~II while both provide an implementation of the {\it method} Gene Set Enrichment Analysis. Important aspects to consider when choosing between an FCS~I and FCS~II tool are provided in Section 2.4 in the supplement.

\begin{table}[ht]
\centering
\caption{Distinction of FCS tools based on the form of the required input object. While FCS~I comprises those FCS tools that accept as input the gene expression data set as a whole, FCS~II tools require a gene ranking that reflects the genes' magnitudes of differential expression between both conditions.}

\label{fcs_distinction}
\begin{tabular}{l|c}
 FCS tool & FCS~I or FCS~II? \\
\hline
\tool{GSEA}  & FCS~I \\
\tool{GSEAPreranked} & FCS~II \\
\tool{clusterProfiler}'s GSEA & FCS~II \\
\tool{PADOG} & FCS~I \\
\hline
\end{tabular}
\end{table}

\subsubsection{Gene Set Enrichment Analysis}    
The method Gene Set Enrichment Analysis \citep{subramanian2005gene} is classified as FCS and starts with a gene ranking that is based on the individual gene's association with both conditions. \\
For a given gene set, the enrichment score is obtained by going down the gene ranking step by step and successively increasing a mathematical term for each gene that is a member of the gene set, while decreasing the term for each gene that is not. Each increase is weighted by the respective gene's association with the conditions, causing genes at the top and bottom of the ranked gene list to contribute more strongly to the enrichment score. The enrichment score is then obtained as the maximum deviation from zero of this term. Note that a more detailed description of the computation of the enrichment score is provided in Section 2.5 in the supplement. By default, significance is assessed by testing a self-contained null hypothesis. \\
For the purpose of clarity, we use the written out term `Gene Set Enrichment Analysis' for the method and the corresponding abbreviation `\tool{GSEA}' for its implementations (i.e. tools).

%%%%%%%%%%%%%%%%%%%%%%%%%%%%%%%%%%%%%%%%%%%%%%%%%%%%
\subsubsection{GSEA}
The methodology behind the method Gene Set Enrichment Analysis (see previous section) is implemented in the web-based tool \tool{GSEA} \citep{subramanian2005gene}. This tool requires the input of the gene expression data set as a whole and is therefore classified as FCS~I. By default, a gene ranking that represents the association with both conditions is generated internally using the signal-to-noise ratio. \\ 
In accordance with the self-contained null hypothesis, the enrichment of a given gene set is evaluated by randomly generating 1000 phenotype permutations.

\subsubsection{GSEAPreranked}
Additionally, the web-based tool \tool{GSEA} (see previous section) offers the version \tool{GSEAPreranked} \citep{subramanian2005gene} for which a user inputs their own list of genes that has already been ranked by a suitable gene-level statistic of choice. This version, which is categorized as FCS~II accordingly, is recommended over the traditional tool \tool{GSEA} if the gene-level statistics provided by the web-based tool do not suit the gene expression data at hand. Further important aspects when choosing between \tool{GSEAPreranked} and \tool{GSEA} can be obtained from Section 2.4 in the supplement. In accordance with FCS~II, GSEAPreranked assesses the significance of a given gene set using gene set permutation. 

\subsubsection{PADOG}
PADOG \citep{tarca2013comparison}, which is short for `Pathway Analysis with Down-weighting of Overlapping Genes', is an FCS method whose implementation \tool{PADOG} is categorized as FCS~I. It assigns a higher weight to those genes in the computation of the enrichment score that are gene set-specific, meaning that they are assigned to a few or even only a single gene set and might therefore indicate a true connection between the associated gene set and the conditions of interest. \\
After quantifying each gene's magnitude of differential expression using a moderated variant of the t-statistic \citep{smyth2004linear}, a gene set's enrichment score is obtained as the weighted mean of the t-statistics of the corresponding gene set members, with each gene's weight depending on its frequency of gene set memberships. Eventually, the significance of a standardized version of the enrichment score is assessed by testing a self-contained null hypothesis through phenotype permutation. \\ It is noted that this method is implemented in the identically named \texttt{R} package \tool{PADOG}. 

\subsubsection{clusterProfiler's GSEA}
The \texttt{R} package \texttt{clusterProfiler} implements a modification of the method Gene Set Enrichment Analysis (see Section `Gene Set Enrichment Analysis'). This implementation requires a list of the genes from the experiment ranked by their magnitudes of differential expression as input and is therefore classified as FCS~II. From this gene ranking, an enrichment score is calculated analogously to the regular methodology of Gene Set Enrichment Analysis. However, the difference to the standard methodology, in which the conditions of the samples are permuted a certain number of times, is that significance is assessed by generating an empirical null distribution of enrichment scores by permuting the gene labels within the ranked list \citep{dose_ref}. Note that this approach also differs from gene set permutation.

%% file: 04_preprocessing.tex
\section{Preprocessing and Practical Aspects}

\begin{figure*}
\centering
%\captionsetup{justification=centering}
       
\includegraphics[scale=0.7]{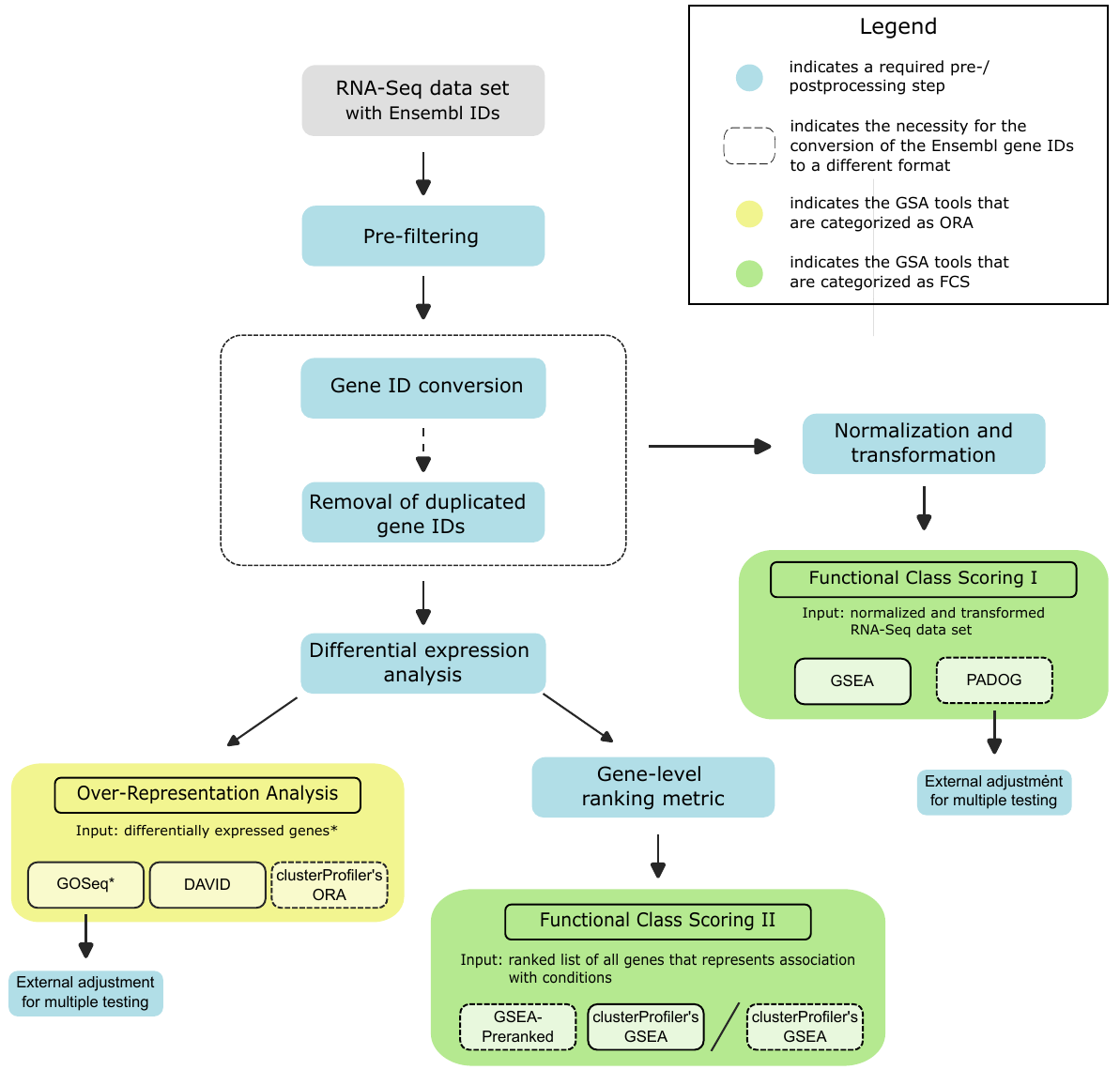}
\caption{\small Overview of the practical GSA workflow for each GSA tool and an RNA-Seq data set with genes identified in the Ensembl ID format. A special focus is placed on the required preprocessing steps. The dashed borders indicate the necessity for a conversion of the gene IDs which is the case when the corresponding GSA tool does not accept the gene IDs in the Ensembl ID format. An overview of the accepted gene ID format(s) per GSA tool is provided in Section 3.2.1 in the supplement. Note that for \texttt{clusterProfiler}'s GSEA tool, the necessity for the conversion of gene IDs depends on the choice of the gene set database (further information is provided in the supplement). `*' indicates that for \tool{GOSeq}, the required input object consists of a named binary vector which indicates differential expression, as opposed to the list of differentially expressed genes. \tool{GOSeq} and \tool{PADOG} are the only two GSA tools from the selection for which the user has to perform multiple testing adjustment manually after running the tool. }\label{fig_pre_workflow}
        \end{figure*}

In general, a user has to procure a high amount of information about the preprocessing steps required to generate the input object for the chosen GSA tool and additional practical aspects that are important when running it. A clear overview of the necessary steps is complicated by a discrepancy in the quality of the instructions and user manuals between the different tools. Furthermore, there is commonly a lack of default procedures among the wide variety of different approaches to serve as guidance for the user. The enormous amount of informal guidance provided in online communities that are based on personal experience, as opposed to evidence-based guidelines from scientific literature, illustrates this issue. \\
Another difficulty a user typically encounters is that the practical steps required to conduct GSA can differ greatly across different methods and tools and particularly between the different approaches.  However, even for tools within the same class of GSA, there are exceptions regarding the form of the required input object. \\
To offer the user a guide through this seemingly obscure procedure of preprocessing, we have summarized the steps required to generate the accepted input object for each of the GSA tools from Table \ref{overview_methods_tools} for an RNA-Seq data set with its genes identified in the Ensembl ID format \citep{cunningham2022ensembl}. Additionally, a graphical illustration is provided in Figure~\ref{fig_pre_workflow}. While we focus on the creation of the input objects for the different GSA tools, we also discuss important practical aspects within the tools, such as the gene set database. Note that in this section, we will (almost) exclusively refer to GSA {\it tools} instead of {\it methods} since preprocessing takes place in the practical context of the tools. In particular, even those tools that offer an implementation of the same method can differ in their accepted input object so the term \textit{method} may be misleading. An extensive review of the researchers' degrees of freedom in the individual steps of preprocessing, leading to data preprocessing uncertainty, is provided in Section 5 in the supplement.

%%%%%%%%%%%%%%%%%%%%%%%%%%%%%%
%%% Figure for preprocessing Steps 
%%%%%%%%%%%%%%%%%%%%%%%%%%%%%%

\subsection{Pre-Filtering}
One of the first steps in the preprocessing of the RNA-Seq data, before GSA as well as differential expression analysis, is pre-filtering which refers to the exclusion of lowly expressed genes. The reasons for this procedure are discussed particularly in the context of differential expression analysis. One reason behind pre-filtering is that lowly expressed genes are unlikely to be detected as differentially expressed from the start and that their exclusion leads to a reduction of the number of statistical tests (which equals one per gene) to be performed \citep{love2014moderated}. As a consequence, the loss of statistical power accompanied by the correction for multiple testing is alleviated. The authors of edgeR \citep{robinson2010edger} argue from a biological viewpoint and state that a certain minimum expression level of a gene is required for it to be of biological importance. \\
In the context of GSA, the exclusion of genes with a particularly low magnitude of counts in RNA-Seq data is substantiated by the high likelihood that these counts are erroneously mapped to the given gene, even though it is not actually expressed in any of the samples \citep{subramanian2005gene}.

\subsection{Gene ID Conversion and Removal of Duplicated Gene IDs}
A variety of formats to identify gene IDs in a gene expression data set exist, such as Ensembl gene IDs \citep{cunningham2022ensembl}, NCBI (Entrez) gene IDs \citep{sayers2022database} and HGNC (HUGO) gene symbols \citep{seal2023genenames}, and oftentimes, there is a discrepancy between the gene ID format in the given gene expression data set and the format(s) accepted by the chosen GSA tool (see Section 3.2.1 in the supplement for an overview of the gene ID formats accepted by the GSA tools). In this case, the gene IDs of the gene expression data set must be converted to the required format. \\
Different conversion tools rarely lead to an identical conversion pattern of the genes and guidance on the optimal choice of a conversion tool among the many options is scarce. Moreover, for some gene IDs in the initial format, the chosen conversion tool may not provide any corresponding gene ID in the converted format. These genes are lost for subsequent analysis and the gene expression data set is reduced accordingly. \\
Furthermore, there is often an ambiguous mapping between two distinct gene ID formats within a single conversion tool which can lead to duplicated gene IDs in two ways. Firstly, a gene ID of the initial format can be converted to several gene IDs in the required format. Secondly, multiple distinct gene IDs of the initial format can be converted to the same gene ID in the required format. The two cases are addressed in detail in Section 3.2.2 in the supplement. \\
Both circumstances require manual handling, however, there is a lack of guidelines on an optimal manner of duplicate gene ID removal provided by scientific publications. More specific details on possible approaches to removing duplicated gene IDs are provided in Section 5.2 in the supplement.

\subsection{Normalization of RNA-Seq data}
In many cases, conducting GSA requires a suitable normalization of the RNA-Seq data set during preprocessing. Normalization aims to remove any sample-specific biases from RNA-Seq data that arise from the sequencing process itself and would, unless accounted for, hinder an accurate comparison between different samples. An overview of the biases that are typically addressed by normalization methods, such as compositionality effects or differences in library size, is provided in Section 3.3 in the supplement. \\

\subsection{Transformation of RNA-Seq Data}
In addition to normalization, a suitable transformation of the RNA-Seq data set is required if the chosen GSA method was specifically developed for data obtained using microarray technology. Note that this technology was the state-of-the-art technology to quantify gene expression before RNA sequencing emerged. For instance, the web-based tool \tool{GSEA} and the \texttt{R} package \texttt{PADOG}, which are classified as FCS~I, use the metrics signal-to-noise ratio and the moderated t-statistic, respectively, to generate the gene ranking. Such metrics typically assume continuous and particularly homoscedastic quantities, i.e. (approximately) equal standard deviations of gene expression measurements across all magnitudes of quantified gene expression. While these assumed characteristics match those of microarray measurements, they do not hold for RNA-Seq data \citep{law2014voom}. To address this issue, several methods have been proposed that aim to make microarray methods applicable to RNA-Seq data by addressing the discreteness and heteroscedasticity of these measurements with a suitable transformation (see Section 5.5 in the supplement). \\
While in the context of differential expression analysis, the transformation of RNA-Seq data is particularly discussed by the authors of the method voom/limma \citep{law2014voom} (see Section 5.6.2 in the supplement), the necessity of such transformation seems to be far less present in the context of GSA. For instance, in the user manual provided for the web-based tool \tool{GSEA}, it is stated that even though \tool{GSEA} is commonly used for RNA-Seq data, its applicability to RNA-Seq measurements has not been fully investigated yet. This implies that \tool{GSEA} is typically run on RNA-Seq data without any transformation to properly align its characteristics to microarray measurements. The only recommendation made is to perform a suitable method for normalization such as provided by DESeq2 \citep{love2014moderated} which on its own is not sufficient to align the characteristics of RNA-Seq data to those of microarray measurements. \\
We note that a transformation of the RNA-Seq data can be bypassed for ORA and FCS~II, whose required input objects are typically generated using differential expression analysis, by choosing a method for differential expression analysis specifically developed for RNA-Seq data (see Section `Differential Expression Analysis').

\subsection{Differential Expression Analysis}
The goal of differential expression analysis is to assess whether individual genes from the experiment exhibit different expression patterns between the two conditions. Typically, from the results of such analysis, the genes with an adjusted $p$-value below a certain threshold are classified as differentially expressed, whereas the remaining genes are categorized as not differentially expressed. As already described in Section `Over-Representation Analysis (ORA)', this dichotomous interpretation of differential expression analysis is used by ORA, which usually requires as input a list of differentially expressed genes. \\ 
In FCS, differential expression analysis is conducted in the course of preprocessing only for those tools that require as input a gene ranking generated by the user, i.e. FCS~II, while this ranking is created internally for FCS~I. In the context of GSA, \citet{reimand2019pathway} recommend using methods for differential expression analysis that employ a variance stabilization, such as limma/voom \citep{law2014voom}, DESeq2 \citep{love2014moderated}, or edgeR \citep{robinson2010edger}.

\subsection{Gene-Level Statistic (for FCS~II)}

For FCS~II tools, the required input list of all genes from the experiment that are ranked by their magnitude of differential expression is generated using a suitable gene-level statistic. This gene-level statistic is typically based on the quantities from the results table of the preceding differential expression analysis, such as the (un-adjusted) $p$-value and the estimated log fold change between both conditions. The gene-level statistic of each gene (and therefore the gene ranking) is generated such that genes that show gene expression strongly in favor of one of both conditions have a high positive or negative value (and are placed at the upper or lower end of the ranking), respectively, while genes whose expression behavior does not change between both conditions have a value near zero and are located in the middle. For further information and suggestions on gene-level statistics, inspect Section 5.7 in the supplement.

\subsection{Adjustment for Multiple Testing}
In the context of GSA, a statistical test is performed for each gene set provided by the given gene set database, resulting in the necessity for multiple testing adjustment (for a more detailed explanation, refer to Section 5.8 in the supplement). While the majority of GSA tools considered in this work perform multiple testing adjustment internally, \tool{GOSeq} and \tool{PADOG} constitute an exception. Both tools require multiple testing adjustment to be performed by the user, which can be done using \texttt{R} package \texttt{base}.

\subsection{Gene Set Database}
As mentioned in Section `General Principle', the gene set database provides information on how the genes from the experiment are aggregated into gene sets based on a specific relationship between the respective genes. Such a relationship could be a common chromosomal location or biological function \citep{subramanian2005gene}. In the context of ORA, the gene set database additionally provides the set of background genes (i.e. universe) if the accepted input object consists of the list of differentially expressed genes and the information about the entirety of genes from the experiment is lost. \\
There are a variety of different gene set databases which typically differ in aspects such as the number of gene sets they contain \citep{mubeen2019impact}. \citet{mubeen2019impact} observe that in practice, the choice of the gene set databases is primarily made based on personal experience and preference, such as a tendency towards gene set databases that empirically yield preferable results, even though it should be made based on the underlying biological context.

\subsection{Specification of a Seed for Reproducibility (for FCS)}
FCS tools, irrespective of the required input object, typically generate a large number of random permutations to empirically calculate a $p$-value of enrichment. To ensure that the same permutations are generated when the tool is run at different points in time and thus lead to identical results, a seed can be specified in the form of an arbitrary integer number. The specification of a seed is therefore highly recommendable. \\
In contrast to FCS, ORA typically does not involve a random component as no random sampling or permuting is performed. Instead, enrichment is assessed based on parametric assumptions on the underlying null distributions. This also applies to the methods for differential expression analysis, resulting in no necessity and therefore no option to specify a seed. 

%% file: 05_implications.tex
\section{Uncertainties, Implications, and Recommendations}

\subsection{Method, Parameter, and Data Preprocessing Uncertainty}
The existence of method uncertainty becomes clear in Section `Methods and Tools', in which a variety of methods and tools are presented. It is important to keep in mind that this selection is only a small subset of the plethora of methods and tools available, a fact that further intensifies method uncertainty. \\
Even within an individual GSA tool, there is typically a certain number of parameters that can be modified to adapt the analysis strategy to the given research question and the data at hand (e.g. the gene set database), resulting in parameter uncertainty. Since the majority of flexible parameters differ between ORA and FCS, we elaborate on the common flexible parameter choices for both approaches separately (Sections 4.1 and 4.2 in the supplement). 
\\
As already described, the flexibility offered between and within the individual GSA methods and tools is accompanied by a variety of options in the preceding preprocessing for which guidance in the form of default approaches is scarce. Furthermore, in existing works on applied research, the exact details of the performed steps of preprocessing are typically neglected, if indicated at all. We address the corresponding data preprocessing uncertainty in the supplement (see Section 5) by providing an overview of existing strategies from published literature for each step from Figure~\ref{fig_pre_workflow}. 

\subsection{Implications and Recommendations for Users}

Awareness of the uncertainties related to the use of GSA methods and corresponding tools is the first step toward their correct use. In particular, the mere name of the method or tool is not sufficient to characterize an analysis. A researcher, for example, cannot assume that they used the same analysis pipeline as used in a previous publication just because they used the same tool. Attention should be devoted to all details of the implementation.

As far as possible, it is generally recommended to choose methods and parameters considering the biological setting and research question in the first place---as opposed to choices made after seeing the results as will be further discussed at the end of this section. 

It is also important to note that, while some parameters can (or must) be chosen by the user to adapt to the underlying biological context and offer a certain amount of flexibility, others are to be considered as `technical' parameters related to the reproducibility of the results and computational resources. The latter thus do not offer such flexibility in the strict sense. In particular, the number of permutations in permutation-based methods should be set as large as computationally feasible. One should not run the analyses with different numbers of permutations but rather set this number to a large value in the first place. The seed---for methods/tools involving a random component---is also not a parameter that should be specified flexibly. The seed should be, if at all, changed only to check whether results obtained with different seeds are very similar. It should particularly never be changed in the hope to obtain `more satisfying results'.

All changes of parameters or options conducted after seeing the results of interest, i.e. a deliberate exploitation of the uncertainties in the GSA workflow, may tempt researchers to selective reporting. Typically, researchers will come back to their initial analysis and results if the change does not make the results more satisfying in their view. In contrast, they will adopt the changed pipeline if they prefer its results to those of the original pipeline. Such practice may seem natural at first glance so that many researchers who proceed this way may not be aware that this excessive fitting of the analysis strategy to the given data set is a form of cherry-picking, also denoted as fishing for significance in the context of statistical tests. In other contexts than GSA, such a behavior has been demonstrated to yield over-optimistic research findings that cannot be validated on new, independent data \citep{ullmann2023over} and thus contribute to the so-called replication crisis. In this context, we want to emphasize that research findings that cannot be replicated on independent data are not valid. In the specific context of GSA, the extent to which over-optimistic results can be produced when method, parameter, and data preprocessing uncertainty are exploited has not yet been explored.

However, one may choose to apply different pipelines in the first place as a form of sensitivity analysis to check the robustness of the results. This amounts to the `report uncertainty strategy' suggested previously as one of many potential approaches to handling the multiplicity of analysis strategies \citep{hoffmann2021multiplicity}. The challenge will then be to present the results in a clearly arranged way to avoid confusion.

Irrespective of whether such robustness checks are performed or not, all choices done by researchers performing GSA should be transparently reported. In this context, code should be made available for the purpose of reproducibility, because all details of an analysis can barely be entirely reported in the form of text. Along with the code, researchers should always indicate the version of the chosen GSA tool and the gene set database. In a recent survey on more than 200 gene set analyses, it was found that only a minority indicates the version of the software used while even less provide information on the version of the gene set database \citep{wijesooriya2022urgent}.

\subsection{Implications and Recommendations for Method Developers}
Considering the possibly important impact of uncertain choices on the results and the devastating effect of cherry-picking in terms of replicability, method developers should as far as possible provide practical guidance regarding parameter and data preprocessing uncertainty to future users. This could be done in the form of defaults and clear instructions regarding deviations from the defaults, which are ideally evidence-based rather than resulting from anecdotal experience.

Regarding the high degree of freedom affecting GSA, we regard neutral comparison studies \citep{boulesteix2017towards} investigating the behavior of methods dependent on parameters and preprocessing in different scenarios as playing a crucial role toward the more proficient use of these methods by users, ultimately leading to more reliable results in the field. In particular, keeping in mind that the `one method fits all data sets' philosophy does not sufficiently reflect the complexity of such analyses \citep{strobl2022against}, we argue that comparison studies should investigate a variety of settings, data sets, and methods' variants in addition to the default parameters. Particularly, factorial comparison designs are used to identify which features of the methods or pipelines make a difference. Suppose as a simplified example of such a factorial comparison design that the authors of a method we call `A' apply it together with a particular metric for differential expression assessment, while the authors of another method called `B' use another metric. In this case, it may make sense to investigate the behavior of all $2\times 2$ combinations of methods and metrics rather than sticking to the original two combinations, as already demonstrated in another context \citep{niessl2023explaining}. \\

%% file: 06_supplementary_material.tex
\textbf{\Large Supplemental Material}

\section{General Principle}
In this section, we supplement the information from the corresponding Section `General Principle' in the main document with additional aspects important for the conduct of GSA. Since this supplement focuses on the researchers' degrees of freedom, we do not cover each subsection from the main document but instead, add supplementary information where it is relevant.

\subsection{Over-Representation Analysis (ORA)}
The null hypothesis implies that the given gene set is the result of a random sampling of $G$ genes from the entirety of genes in the experiment, namely $N$. Consequently, the probability distribution of $H$ differentially expressed genes within the given gene set can be modeled using the hypergeometric distribution \citep{draghici2003global}: 
\begin{align}
\label{hypergeometric_distribution}
    f(H;N,G,L)= \frac{\binom{G}{H} \cdot \binom{N-G}{L-H}}{\binom{N}{L}}.
\end{align}
This corresponds to Fisher's exact test and the $p$-value of over-representation is consequently calculated as 
\begin{align}
\label{fisher_exact}
    P&=\sum_{j=H}^{G} f(j;N,G,L)=1-\sum_{j=0}^{H-1} f(j;N,G,L).
\end{align}
In the case of a large number of genes in the expression data set, which is typical for high-throughput experiments, the calculation of the hypergeometric distribution tends to be computationally extensive. Consequently, an approximation with the Binomial distribution is used to calculate the $p$-values of over-representation \citep{draghici2003global}. Eventually, inference is based on adjusted $p$-values resulting from multiple testing adjustment since a $p$-value of over-representation is calculated for each gene set in the experiment.

\subsection{Pathway Topology (PT)}
Methods assigned to PT proceed similarly to FCS methods with the major difference that they additionally model interactions between the genes \citep{reimand2019pathway}. This difference manifests in the calculation of the gene-level statistic for each gene from the experiment \citep{khatri2012ten}.

\section{Methods and Tools}
In this section, we supplement the information from Section `Methods and Tools' in the main document with additional aspects that are relevant or, more importantly, with aspects that can be modified by the user. Note that the general focus in the supplement is placed on the researchers' degrees of freedom so that we do not address each GSA method or tool in detail. 

\subsection{Selection Criterion}
As described in the main document, the selection of GSA methods and tools considered in this paper is based on the work of \citet{xie2021popularity} who provide a database of existing GSA methods and tools together with a ranking on the overall popularity, which is based on the number of citations (up to September 2019), and on recent popularity derived from the number of citations from May 2018 to April 2019. From the former ranking, we have selected the top two tools \tool{GSEA} \citep{subramanian2005gene} (including its variant \tool{GSEAPreranked}) and \tool{DAVID} \citep{huang2009bioinformatics, huang2009systematic}. While \tool{GSEA} and \tool{GSEAPreranked} are a direct and a modified implementation of the method Gene Set Enrichment Analysis, respectively, \tool{DAVID} carries out a variant of the general methodology of ORA. Additionally, from the ranking on recent popularity, we have extended this selection by GOSeq \citep{young2010gene} and the \texttt{R} package \texttt{clusterProfiler} \citep{wu2021clusterprofiler}, which offers an implementation to ORA and Gene Set Enrichment Analysis. In contrast, the method PADOG \citep{tarca2013comparison} has been selected as a method with fewer citations but that is listed as one of only a few methods with a good performance in more than one simulation study considered by \citet{xie2021popularity}. Note that we consider GOSeq and PADOG as two methods that are implemented in the identically named \texttt{R} packages \tool{GOSeq} and \tool{PADOG}, respectively.

\subsection{DAVID}
DAVID starts from Table 1 from the main document in which the frequencies of differential expression (yes/no) are displayed against the membership of the given gene set (yes/no) for all genes from the experiment. Then, in this contingency table, one (arbitrary) gene that is a member of the given gene set is moved from the list of differentially expressed genes to the set of remaining genes, resulting in the contingency table in Table \ref{contingencytable_david}. More specifically, quantity $H$ becomes $H-1$ whereas quantity $G-H$ changes to $G-H+1$, accordingly. \\
Analogous to the regular ORA framework, a $p$-value (called EASE score) is then obtained from the hypergeometric distribution and Fisher's exact test. However, the resulting $p$-values are higher compared to those resulting from regular ORA. While this increase is small for larger gene sets, it is considerably higher for gene sets that contain only a few genes, which is the intention behind this modification of the initial contingency table. In the special case of a gene set containing only a single gene, the EASE score amounts to 1, i.e. differential enrichment is precluded completely.

\begin{table}[h]
\centering
\caption{Contingency Table in DAVID generated from all genes in the experiment. N: Total number of genes from the experiment; H: Number of genes from input that are members of the given gene set (`hits'); L: number of genes in input list, G: number of genes that are a member of given gene set}
\label{contingencytable_david}
\begin{tabularx}{0.8\textwidth}{p{0.2\textwidth} |Y{0.2\textwidth} Y{0.2\textwidth}|p{0.2\textwidth}}
 & Differentially expressed & Not differentially  expressed & total \\
\hline
Member of \newline gene set & $H-1$ & $G - H +1$& $G$ \\
Not a member \newline  of gene set & $L-H$ & $N-L-(G-H)$ &$N-G$ \\
\hline
total & $L-1$ & $N-L+1$& $N$\\
\end{tabularx}
\end{table}

\subsection{Length Bias in GOSeq}\label{lengthbias_goseq}
The so-called `length bias', also called `transcript length bias' or `selection bias', is inherent to RNA-Seq data and a consequence of the mRNA fragmentation in the sequencing process described in Section \ref{rna_seq} \citep{oshlack2009transcript}. Since gene expression of a given gene is measured by the number of snippets resulting from the fragmentation of its mRNA strands in a given sample, genes with longer transcript length (i.e. length of the mRNA strands) are automatically assigned a higher number of read counts, independent of their actual expression level. Since for count data, statistical power increases with the magnitude of counts, genes with longer transcript length are more likely to be detected as differentially expressed compared to shorter genes, even if their respective expression levels are equal. \\
For information on how length bias is addressed in GOSeq, return to Section `GOSeq' in the main document.

\subsection{Distinction: FCS~I vs. FCS~II}
There are several aspects to consider when choosing between FCS~I and FCS~II tools. First, gene set permutation disregards likely interactions between genes that are members of the same gene set. This can cause false detection of the gene sets as differentially enriched solely due to high correlations between the genes \citep{maleki2020gene}. Consequently, the results may indicate a high number of differentially enriched gene sets, among which a high proportion are false positives. This difficulty is bypassed in phenotype permutation since the composition of the gene sets in terms of genes is fixed. \\
Second, in the course of phenotype permutation, a certain number of permutations of the sample conditions are performed to generate a null distribution of enrichment scores. This number is typically high (such as 1000 by default in \tool{GSEA} and \tool{PADOG}), therefore requiring a relatively high sample size. For gene expression data sets of smaller sample sizes (such as two to five samples in each condition \citep{reimand2019pathway}), phenotype permutation may therefore not be applicable and the user has to resort to gene set permutation. This is achieved by either the choice of an FCS~II tool or by switching to gene set permutation in an FCS~I tool (if possible). \\
We note that some FCS tools perform phenotype permutation by default and are therefore classified as FCS~I, however, the user sometimes has the choice to switch to gene set permutation.

\subsection{Gene Set Enrichment Analysis}
\label{GSEA_theory}

Gene Set Enrichment Analysis is an FCS method that was proposed by \citet{subramanian2005gene}. 
The starting point of Gene Set Enrichment Analysis is a ranked list $L_{r}$ which contains each of the $N$ genes whose expression was measured in the experiment. In accordance with Section `Functional Class Scoring (FCS)' in the main document, this ranked list is generated from a suitable gene-level statistic and represents the extent to which each gene is differentially expressed between both conditions. Correspondingly, a position at the top or bottom of the list indicates that the gene expression pattern is strongly favored towards the first and second phenotype, respectively. 
%%%%%%%%%%%%%%
The general idea of Gene Set Enrichment Analysis is to investigate whether the members of a given gene set $S$ are distributed more towards the top or bottom of $L_{r}$, which indicates differential enrichment, or else spread across it randomly. In this context, the overall framework consists of the following three steps: \\
%%%%%%%%%%%%%%
In the first step, an enrichment score (ES) is calculated for each gene set which represents the extent to which it is represented at the top or bottom of $L_{r}$. To describe the process in mathematical terms, the value of the gene-level statistic of gene $g_{i}$ is denoted by $r_{i}=r(g_{i})$. The genes are then ranked with respect to $r_{i}$ in a decreasing manner such that the gene ranking has the form $L_{r}=\{g_{1},...,g_{N}\}$. In particular, gene $g_{1}$ shows the strongest gene expression towards the first condition whereas $g_{N}$ shows the strongest gene expression towards the second condition. For a gene set $S$ with $G$ genes, the enrichment score is then calculated by running down the ranked list $L_{r}$ and evaluating the following two sums with each step $l$, $l=1,...,N$: 

\begin{align}
\label{enrichmentscore}
\begin{split}
    P_{\text{hit}}(S,l)&=\sum_{ \substack{g_{i} \in S \\ i \leq l}} \frac{|r_{i}|^{p}}{N_{R}}, \qquad \text{where } N_{R}=\sum_{\substack{g_{i} \in S}} |r_{i}|^{p} \\
 P_{\text{miss}}(S,l)&=\sum_{\substack{g_{i} \centernot \in S \\i \leq l}} \frac{1}{N-G}.
 \end{split}
\end{align}

The term $P_{\text{hit}}(S,l)$ increases for each gene in the gene ranking that is a member of the gene set S. For each gene in the gene set, the amount of the increase, i.e. its weight, is determined by its absolute value of the ranking metric and the exponent $p$. Parameter $p$ specifies how strongly each gene's value of the gene-level statistic contributes to this sum. For instance, for $p=0$, each gene's contribution to the enrichment score is equal and the enrichment score corresponds to a standard Kolmogorov-Smirnov statistic. In contrast to that, the choice of $p=1$, which is the common default, results in those genes with a higher positive or negative absolute value of the gene-level statistic affecting the enrichment score more strongly (weighted Kolmogorov-Smirnov statistic). 
%%%%%%%%%%%%%%%%%
The enrichment score of gene set $S$ is computed as the maximum deviation of
$P_{\text{hit}}(S,l)-P_{\text{miss}}(S,l)$ from $0$. The intuition behind this statistic is that a gene set whose genes are concentrated towards the top or bottom of $L_{r}$ has a high positive or negative enrichment score, respectively, whereas gene sets with randomly spread genes result in a lower absolute enrichment score. \\
%%%%%%%%%%%%%%
The second step focuses on assessing the significance of each gene set's enrichment score. As the default null hypothesis is the self-contained null hypothesis, significance analysis of a gene set $S$ is evaluated using phenotype permutation.
%%%%%%%%%%%%%%%%%
In the third step, a gene set's enrichment score is normalized for gene set size, and its $p$-value is adjusted for multiple testing. Normalization for gene set size is substantiated in the circumstance that the `raw' enrichment score is biased by the size of the gene set, meaning that gene sets that differ in size can systematically differ in their enrichment scores, independent of the gene expression activities of their respective gene set members. Accordingly, the normalized enrichment score of a gene set is obtained by dividing the raw enrichment score by the mean of the enrichment scores obtained from the $1000$ random permutations resulting from the above phenotype permutation.

The resulting normalized enrichment score, abbreviated with `NES', then builds the base for the interpretation of the results. In the last step, significance of a given gene set is evaluated based on the $p$-value which is adjusted for multiple testing, as the overall procedure is performed for a multitude of gene sets.

\section{Preprocessing}

\subsection{RNA-Seq data set} \label{rna_seq}
In this paper, we assume that the gene expression data is generated from the state-of-the-art RNA Sequencing (`RNA-Seq') Technology which has replaced Microarray Analysis Technique as gene expression profiling in the years since its release. In the process of an RNA-Seq experiment, an mRNA strand, which is a gene expression product, is essentially fragmented into shorter snippets that are then, after further processing, sequenced by the high-throughput sequencing technology, resulting in short sequence reads \citep{wang2009rna}. By performing this process with an entire sample of mRNA strands, the level of gene expression of a given gene is assessed as the number of sequence reads that can be assigned to the corresponding gene in this particular sample. Additionally, each sample is assigned a label that corresponds to the condition or phenotype of interest. We here assume that the labels are binary (such as treatment versus control or healthy versus diseased). \\

\subsection{Gene ID Conversion and Duplicated Gene IDs}

\subsubsection{Gene ID Conversion}

The GSA tools considered in this work can differ greatly in the selection of gene ID formats they accept, so that the user has to inspect the corresponding user manual of the chosen method carefully. Furthermore, for some tools, it is more difficult to get explicit information on the accepted gene ID formats compared to other tools. To overcome this issue, we provide an overview of the accepted gene ID formats in Table \ref{accepted_geneIDs}. Note that if the GSA tool accepts the genes in the format they are initially identified in the gene expression data set, it is advisable to work with exactly this format. The reason for this is that gene ID conversion leads to a loss of those genes for which no corresponding gene ID in the converted format exists.

\begin{table}[ht]
\small
\centering
\caption{Overview of the gene ID formats accepted by the GSA tools considered in this paper}

\label{accepted_geneIDs}
\begin{tabularx}{0.95\textwidth}{l|l}
\hline
 GSA Tool &  Accepted gene ID formats \\
\hline
\tool{DAVID} & Wide variety of formats, including \\
& \tabitem Ensembl gene ID and \\ 
& \tabitem NCBI (Entrez) gene ID \\
\tool{GOSeq} & \tabitem Ensembl gene ID \\
& \tabitem NCBI (Entrez) gene ID \\
& \tabitem HGNC gene symbol \\
\tool{clusterProfiler}'s ORA & \tabitem NCBI (Entrez) gene ID \\
\tool{clusterProfiler}'s GSEA & Depends on the choice of the gene set database: \\
& \textbf{for GO}: Variety of formats, such as \\
& \tabitem Ensembl gene ID \\
& \tabitem NCBI (Entrez) gene ID \\
& \textbf{for KEGG}: \tabitem NCBI (Entrez) gene ID \\
\tool{GSEA} & Wide variety of gene ID formats: \\
& \tabitem Ensembl gene ID \\
& \tabitem NCBI (Entrez) gene ID \\
& \tabitem HGNC gene symbol \\
\tool{GSEAPreranked} & \tabitem Recommended: HGNC gene symbol \\
\tool{PADOG} & \tabitem NCBI (Entrez) gene ID\\
\hline
\end{tabularx}
\end{table}
\tool{DAVID} and the web application \tool{GSEA} accept a wider variety of gene identifiers compared to the other tools in this selection. They can especially convert the gene IDs and handle the resulting duplicates internally. \tool{GSEA} typically converts the initial gene IDs to HGNC gene symbols \citep{seal2023genenames} based on which the analysis is then performed. This converting approach is based on the gene expression data set as a whole, resulting in an important distinction between the regular tool \tool{GSEA} and its variant \tool{GSEAPreranked}. Since for \tool{GSEAPreranked}, an already ranked list of the genes created outside of the tool must be provided, the actual gene expression data is not available. Therefore, the user is recommended to convert the initially given gene IDs to HGNC gene symbols before running \tool{GSEAPreranked} and to provide the ranked gene list with HGNC gene symbols as the gene identifiers. \\
Within \tool{clusterProfiler}, the accepted gene ID formats are not consistent. While for the ORA tool, the only accepted gene ID format is NCBI (Entrez) gene ID, it differs between the gene set database GO and KEGG when using \tool{clusterProfiler}'s GSEA tool. For KEGG, the user is, again, restricted to the NCBI (Entrez) gene ID format. For GO, however, the user has the choice between a variety of formats, among which are the Ensembl and NCBI (Entrez) gene IDs. \\
While for \tool{PADOG}, the accepted gene ID formats are restricted to NCBI (Entrez) gene IDs, the user has the choice between Ensembl and NCBI (Entrez) gene IDs as well as HGNC gene symbols in \tool{GOSeq}.

\subsubsection{Duplicated Gene IDs}

\begin{figure*}
\centering
\includegraphics[scale=0.8]{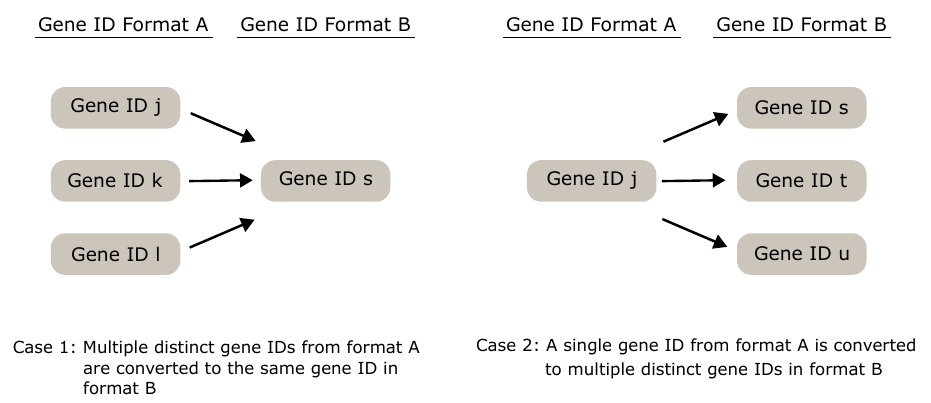}
\caption{Overview of the two ways in which a duplication of the gene IDs can occur as a consequence of conversion from format A (such as Ensembl gene ID) to format B (such as NCBI (Entrez) gene ID).}\label{ID_duplicates}
        \end{figure*}

If in a gene expression data set, the genes are identified in format A but the GSA tool at hand accepts the gene IDs to be in format B, there are two ways in which gene ID conversion can lead to duplicated gene IDs. Firstly, there might be several distinct gene IDs from format A that are converted to the same gene ID in format B. In this case, a conversion results in gene expression data in which several rows with different gene expression data are identified with the same gene ID (see case 1 in Figure \ref{ID_duplicates}). Analogously, there might be a single gene ID in format A which is converted to several distinct gene IDs in format B. In this case, the gene expression data set resulting from conversion contains several rows with identical gene expression data but different gene IDs (see case 2 in Figure \ref{ID_duplicates}). Both of these ways in which duplicated gene IDs can occur must be handled so that each gene ID and each row of count data occur exactly once in the gene expression data set. In section \ref{duplicateID_removal}, we review three approaches to the removal of the duplicated gene IDs.

\subsection{Normalization}

The first technical bias present when comparing samples of RNA-seq data is the library size (also called `sequencing depth'). It corresponds to the total number of read counts of a given sample and results from the sequencing machine, leading to different samples having different library sizes independent of the level of gene expression. The library size therefore does not offer any information on actual expression levels and a difference in library sizes between samples prevents any comparison \citep{love2014moderated}. Linked to this relation is the effect of compositionality which refers to the circumstance that RNA-Seq measurements within a given sample only provide information of relative abundance. For instance, if the majority of counts in a given sample are mapped to a single gene, other genes in this sample will appear to be less expressed in comparison to other samples and therefore may be erroneously detected as down-regulated \citep{robinson2010scaling}. \\ 
Another factor that is addressed by some normalization methods is the gene length since genes with a longer transcript length are naturally assigned more reads in the sequencing process. In the context of differential expression analysis, comparisons are made between different samples, but not between genes within a sample. As gene length is a fixed quantity for a given gene across the samples, it therefore does not have to be considered by normalization techniques during differential expression analysis. For a comparison of genes within a given sample, on the other hand, normalization for gene length becomes necessary. \\
Note that there are different approaches to the exact incorporation of normalization into the analysis. As already described in the main document, many methods for differential expression analysis perform normalization internally. For DESeq2 and edgeR (see Sections `DESeq2' and `edgeR', respectively), on the one hand, normalization leads to an individual normalization factor for each sample, while the count data remain unchanged. For limma, on the other hand, the cpm-transformed count data are themselves normalized (see Section `voom'). The same applies to RNA-Seq data which has been transformed to align its characteristics to that of microarray measurements using one of both approaches described in Sections \ref{voom_transformation} and \ref{vst_transformation}.

\subsection{Differential Expression Analysis}

Methods for differential expression analysis are distinguished by their parametric or non-parametric nature. While parametric methods make assumptions on the distribution of the underlying data and are therefore divided into methods for microarray and for RNA-Seq data, non-parametric methods can be used irrespective of the form of the underlying data. Parametric methods for RNA-Seq data, among which the most common are voom/limma, DESeq2, and edgeR, are generally more popular than non-parametric methods such as SAMSeq \citep{samseq2013} or NOISeq \citep{tarazona2015data,tarazona2011differential}. Note that in this paper, we focus on parametric methods for differential expression analysis. For a brief description of voom/limma, DESeq2, and edgeR, see Section \ref{uncertainty_deAnalysis}.
\section{Parameter Uncertainty in GSA Methods/Tools}
Note that we deliberately restrict ourselves to the set of choices and corresponding options within a given choice that are appropriate in the underlying statistical or biological setting. In contrast, those parameter choices that would indicate a willful manipulation of the results are, on the other hand, excluded from this overview. Throughout this section, we refer to the respective tools as opposed to the corresponding methods of GSA.

\subsection{Parameter Uncertainty in ORA}

\begingroup
% manipulate horizontal spacing
\setlength{\tabcolsep}{7pt} % Default value: 6pt
% manipulate vertical Spacing
\renewcommand{\arraystretch}{1.5} % Default value: 1

\begin{small}
\begin{table*}[b!]

    \centering
    \caption{Parameter uncertainty: overview of flexible parameters in ORA Tools}
    \label{flexible_adaptions_ORA}
   \renewcommand{\arraystretch}{1} % Default value: 1
    \begin{tabular}
    {p{0.07\linewidth} p{0.43\linewidth} Y{0.02\linewidth}| Y{0.02\linewidth} Y{0.09\linewidth}Y{0.09\linewidth}Y{0.09\linewidth}}

    \toprule
    \multicolumn{2}{c}{Parameters} &\multicolumn{1}{c}{} & &\multicolumn{3}{c}{ORA Tool}\\
    \addlinespace[0.2cm]
    \cline{1-2}\cline{5-7}%
     \addlinespace[0.2cm]

    \multicolumn{1}{c}{Parameters} & \multicolumn{1}{c}{Does the user have the choice ...} & \multicolumn{1}{c}{}& & \tool{DAVID} & \texttt{GOSeq} & \texttt{cluster-} \texttt{Profiler's} ORA \\
    \midrule
     Gene set database & 1. between different gene set databases directly offered by the tool, especially between 
GO (with subontologies Biological Process, Molecular Function, and Cellular Component) and KEGG? &&& \newcrossmark & \newcrossmark & \newcrossmark \\

& 2. to upload a user-defined gene set database? &&& & & \newcrossmark \\

    Universe & to upload an alternative universe, i.e. the set of genes considered to be the entirety of genes from the experiment? &&& \newcrossmark & \newcrossmark & \newcrossmark \\
    Method & to adapt the method to calculate the $p$-value of enrichment? && & & \newcrossmark & \\
    Bias & to specify additional factors (`biases') to correct for in the analysis? &&  & & \newcrossmark & \\

     \bottomrule
    \end{tabular}
\end{table*}
\end{small}

\endgroup

Table \ref{flexible_adaptions_ORA} illustrates that \tool{GOSeq} offers the highest flexibility in the parameter choices among the ORA methods. In particular, there are two parameter choices that are specific to \tool{GOSeq} (see Sections \ref{uncertainty_bias} and \ref{uncertainty_methods}). 

\subsubsection{Gene Set Database}
A parameter that can be chosen flexibly across all ORA tools is the gene set database. While in \tool{GOSeq} and \tool{clusterProfiler}'s ORA, the choice between internally provided gene set databases is restricted between GO (with its subontologies Biological Process, Molecular Function, and Cellular Component) and KEGG, the user can choose from a wide variety of gene set databases in \tool{DAVID}. However, \texttt{clusterProfiler}'s ORA is the only tool in this selection that offers the user to upload a user-defined gene set database.

\subsubsection{Universe (set of background genes)}
The second parameter that can be adjusted in each of the presented ORA tools is the universe, i.e. the set of genes considered to be the entirety of genes from the experiment. For ORA, the size of the universe typically determines the population size of the hypergeometric distribution. As mentioned in the introduction to ORA, over-representation of a given gene set is assessed by comparing its number of differentially expressed gene set members to values modeled from the hypergeometric distribution. It is therefore natural that the population size of the hypergeometric distribution, i.e. the set of background genes in the context of ORA, has a considerable impact on which gene sets are detected as differentially enriched. But, since the input to ORA corresponds to a subset of all genes from the experiment, the information on the entirety of genes, i.e. the set of background genes, is not available. The information on the entirety of genes in the experiment is, however, required to set up the contingency table and the corresponding population size of the hypergeometric distribution. Therefore, applications of ORA typically make assumptions on the set of background genes based on information provided by the gene set database. 

The choice of the universe should therefore be carefully considered. Note that the options to modify the universe differ between \tool{GOSeq} and the remaining two ORA tools considered in this paper. This distinction is substantiated in the fact that while for \tool{DAVID} and \texttt{clusterProfiler}'s ORA, the input corresponds to the list of differentially expressed genes, it consists of a named binary vector of all genes from the experiment for \tool{GOSeq}, which indicates the status of differential expression. Therefore, for \tool{GOSeq}, the information on the entirety of genes from the experiment is, in contrast to the other two ORA tools, available. \\
When the required input list corresponds to a list of differentially expressed genes, then recommendations on the optimal choice of the universe made by published literature are not unambiguous. \citet{tipney2010introduction} recommend including those genes into the set of background genes that have some chance of being detected as differentially expressed. \citet{yon2008use}, on the other hand, are more explicit and advocate for a universe that consists of all genes whose expression is measured in the experiment. In contrast, \citet{geistlinger2021toward} contradict this recommendation with the observation that not all genes from the experiment are assigned to a gene set in the given gene set database. Genes that are no members of any gene set can never be drawn in the context of the hypergeometric distribution, while genes that are members of multiple gene sets have an increased probability of being drawn. On the other hand, they state that choosing the set of genes provided by the gene set database as background has the difficulty that those genes from the gene set database that are not measured in the experiment cannot be drawn in the context of the hypergeometric distribution, either. As an escape from this problem, the authors recommend setting the intersection of all genes from the experiment and the entirety of genes annotated to the given gene set database as background. Note that from a practical context, it is not always straightforward to access the list of genes that are annotated in the given gene set database.\\
As mentioned above, the option to modify the universe for \tool{GOSeq} is different due to the different shape of the required input object. From the entirety of genes whose expression was measured in the experiment (which is directly available from the input object), some of the genes cannot be annotated to any of the gene sets from the chosen gene set database, i.e. there are some genes that are no members of any of the given gene sets. By default, these genes are ignored in the calculation of the $p$-value. However, the user has the option to include them. In the context of the Wallenius' non-central hypergeometric distribution, these genes then correspond to the set of genes outside of the given gene set. Note that the inclusion of the genes that are no members of any gene set was the default in earlier versions of \tool{GOSeq}.

\subsubsection{Bias (specific to GOSeq)} \label{uncertainty_bias}
The flexibility in the remaining two parameters is specific to \tool{GOSeq}. Firstly, the user can specify which bias to account for in the analysis which is by default set to length bias. This option is described in Section \ref{lengthbias_goseq}. Alternatively, the user can adjust for the total number of counts in the expression data instead of only gene length. This adaption facilitates an additional adjustment for a gene's expression level, whereby a higher expression level naturally leads to a higher magnitude of counts and therefore increased statistical power. By accounting for a gene's overall expression level, a user can account for all biases that might influence the statistical power to detect a gene as differentially expressed. This way, a higher weight is assigned to genes with lower counts \citep{young2010gene}. However, the authors state that a possible downside of this adaption is that accounting for the total number of read counts may lead to the removal of the bias resulting from actual differential expression.

\subsubsection{Method for the calculation of the p-value (specific to GOSeq)} \label{uncertainty_methods}
The second parameter whose flexibility is specific to \tool{GOSeq} is the method to calculate the $p$-value of enrichment. In addition to the Wallenius approximation, which is used by default, resampling and the standard hypergeometric distribution are offered as an option to evaluate differential enrichment. In \tool{GOSeq}, resampling is performed by randomly selecting a list of genes to be set as differentially expressed of the same size as the actual set of differentially expressed genes. Note that each gene's probability of being selected is weighted by its corresponding value of the probability weighting function. This process is repeated a large number of times (2000 by default) and with each iteration, the number of genes from the corresponding resampled list of differentially expressed genes that are members of the given gene set are reported. Eventually, a $p$-value is obtained as the fraction of resampling iterations in which at least as many of the genes (selected to be differentially expressed) are members of the given gene set as the actual set of differentially expressed genes. \\
Note that the user is discouraged from this resampling strategy as it is computationally expensive. Furthermore, the user is explicitly advised against the use of the standard hypergeometric distribution as it ignores any biases that might be present in the data.

\subsection{Parameter Uncertainty in FCS} \label{sec_flexibility}

\begingroup
% manipulate horizontal spacing
\setlength{\tabcolsep}{7pt} % Default value: 6pt
% manipulate vertical Spacing
\renewcommand{\arraystretch}{1.5} % Default value: 1

\begin{table}[h]
%\small

    \centering
    \caption{Parameter uncertainty: overview of flexible parameters in the FCS tools. Note that the asterisk `*' denotes that, while the corresponding parameter cannot be chosen flexibly within the tool, it can be specified flexibly as part of the preceding preprocessing. }
     \label{flexible_adaptions_FCS}
   % \resizebox{\textwidth}{!}{%
   \renewcommand{\arraystretch}{1} % Default value: 1
    \begin{tabular}{p{0.1\linewidth} p{0.39\linewidth}Y{0.01\linewidth}|Y{0.01\linewidth} Y{0.065\linewidth}Y{0.065\linewidth}Y{0.065\linewidth}Y{0.065\linewidth}}

    \toprule
    \multicolumn{2}{c}{Parameters} & \multicolumn{1}{c}{}&& \multicolumn{4}{c}{FCS Tool}\\

     \addlinespace[0.2cm]
    \cline{1-2}\cline{5-8}%
     \addlinespace[0.2cm]

    \multicolumn{1}{c}{Parameters} &  \multicolumn{1}{c}{Does the user have the choice ...} &\multicolumn{1}{c}{}&& \tool{GSEA} & \tool{GSEA}-\tool{Preranked} & \tool{PADOG} & \tool{cluster-} \tool{Profiler}'s GSEA \\
    \midrule
    Gene set database & 1. between different gene set databases directly offered by the tool, especially between 
GO (with subontologies Biological Process, Molecular Function, and Cellular Component) and KEGG? && & \newcrossmark & \newcrossmark & & \newcrossmark \\

& 2. to create and upload a user-defined gene set database? &&& \newcrossmark & \newcrossmark & \newcrossmark & \newcrossmark \\

    Gene-Level Statistic & between different metrics to rank the genes? && & \newcrossmark & * & & * \\
    Weight & to specify how strongly each gene is weighted by its correlation with the phenotypes in the calculation of the enrichment score? && & \newcrossmark & \newcrossmark & & \newcrossmark \\

     \bottomrule
    \end{tabular}
\end{table}

\endgroup

Among the considered selection of FCS tools, \tool{PADOG} offers much less flexibility in the underlying parameters than the remaining tools. \\

\subsubsection{Gene Set Database}
While all FCS tools offer the user to upload a user-defined gene set database, \tool{PADOG} is the only tool that does not offer a choice in the internally provided gene set databases. In particular, unless the user uploads a customized gene set database, the choice is automatically restricted to KEGG. In contrast, the web-based tool \tool{GSEA} (and correspondingly \tool{GSEAPreranked}) offers nine collections of gene set databases as part of the Molecular Signature Database (`MSigDB') \citep{liberzon2015molecular}, among which GO and KEGG can be found as well. In \tool{clusterProfiler}'s GSEA, the choice of internally provided gene set databases is between GO (with its subontologies Biological Process, Molecular Function, and Cellular Component), and KEGG. \\

%The entirety of gene set databases that can be selected in the web-based tool \tool{GSEA} is part of the Molecular Signature Database (\lq\lq MSigDB'') \citep{liberzon2015molecular}, which is one of the largest and most popular repositories of gene sets. MSigDB is divided into 9 collections of gene set databases. In this context, KEGG is contained by the collection of \lq\lq curated gene sets'', whereas Gene Ontology, with its subontologies, is part of the \lq\lq ontology gene sets''. The default gene set database in the GSEA web application is the Hallmark Gene Set Collection. This collection consists of 50 gene sets that were generated in a way to reduce redundancy across and heteroskedasticity within gene sets by condensing over 4000 gene sets from MSigDB \citep{liberzon2015molecular}. \\

\subsubsection{Gene-Level Statistic (specific to web-based tool GSEA)}

While for \tool{PADOG}, the user cannot modify the metric to capture each gene's magnitude of differential expression between both conditions, for \tool{GSEAPreranked} and \texttt{clusterProfiler}'s GSEA tool, the associated gene ranking must be generated externally and provided as an input. Accordingly, there is no corresponding flexibility within \tool{GSEAPreranked} and \texttt{clusterProfiler}'s GSEA, even though the user has many options in the creation of the input. With that in mind, the corresponding entries in Table \ref{flexible_adaptions_FCS} for \tool{GSEAPreranked} and \texttt{clusterProfiler}'s GSEA are marked with `*'. \\
In contrast to these three FCS tools, the user can modify the gene-level ranking metric in \tool{GSEA}. Note that, after applying one of the transformations in Section \ref{subsec_transformation}, the transformed expression data are approximately on the $\text{log}_{2}$-scale. In combination with a discrete phenotype, i.e. two binary conditions, this results in three options of gene-level ranking metrics. In all of them, a larger absolute value of the statistic indicates a more distinct expression of the gene between both conditions. \\
The default gene-level statistic to measure a gene's correlation with the phenotype of interest is the signal-to-noise-ratio
\begin{align}
\label{signal2noiseratio}
    \text{signal-to-noise-ratio}= \frac{\bar{\tilde{K}}^{\text{norm}}_{i;0}-\bar{\tilde{K}}^{\text{norm}}_{i;1}}{\sigma_{i;0}+\sigma_{i;1}},
\end{align}
where $\bar{\tilde{K}}^{\text{norm}}_{i;0}$ and $\bar{\tilde{K}}^{\text{norm}}_{i;1}$ are the mean transformed and normalized counts of gene $i$ in the samples ascribed to conditions $0$ and $1$, respectively. Similarly, $\sigma_{i;0}$ and $\sigma_{i;1}$ refer to the standard deviations of the normalized counts in both conditions. \\
In addition to the signal-to-noise-ratio, \tool{GSEA} offers the t-statistic which incorporates the number of samples $m_{0}$ and $m_{1}$ of both conditions (which are constant across all genes). It is, as commonly used, defined as 
\begin{align}
    t=\frac{\bar{\tilde{K}}^{\text{norm}}_{i;0}-\bar{\tilde{K}}^{\text{norm}}_{i;1}}{\sqrt{\frac{\sigma_{i;0}^2}{m_{0}}+\frac{\sigma_{i;1}^2}{m_{1}}}},
\end{align} 
where $\bar{\tilde{K}}^{\text{norm}}_{i;0}$ and $\bar{\tilde{K}}^{\text{norm}}_{i;1}$ are the means of the transformed and normalized counts of gene $i$ in the samples ascribed to conditions $0$ and $1$, respectively.\\
%%%%%%%%%%%%%%%%%%%%
Furthermore, the user can generate the ranking of the genes based on the average fold change between both phenotypes. Due to the transformation of the expression data using voom or DESeq2's varianceStabilizingTransformation, the values are consequently on the log scale so that another suitable ranking metric is Difference of Classes (DoC), namely 
\begin{align}
    \text{DoC}=\bar{\tilde{K}}_{i;0}^{\text{norm}}-\bar{\tilde{K}}_{i;1}^{\text{norm}}.
\end{align}
It is noted that \tool{GSEA} offers additional gene-level statistics, however, these are only applicable for count data on the natural scale or in the case of a continuous phenotype.

\subsubsection{Weight}
For all implementations of \tool{GSEA}, may it be the regular web-based tool, \tool{GSEAPreranked}, or \tool{clusterProfiler}'s GSEA, the user is flexible in the choice of the strength of the weight assigned to the association of each gene with both conditions in the calculation of the enrichment score. In all three implementations, this weight is addressed in the context of the exponent of the absolute value of the gene-level statistic (see exponent $p$ in Equation \ref{enrichmentscore}). A larger exponent results in a higher weight of the genes with higher statistics, i.e. genes that are strongly associated with one of the two conditions, compared to the remaining genes. In contrast, the choice of the exponent value $p=0$ leads to equal weights for all genes in the gene ranking. Note that the default exponent amounts to $p=1$ in all three tools. While in \tool{GSEA} and correspondingly \tool{GSEAPreranked}, the allowed alternative options for the exponent consist of the values $0,1.5,$ and $2$, there is no such indication offered in \texttt{clusterProfiler}'s manual. \citet{subramanian2005gene} suggest a choice of $p < 1$ if the user wants to focus on gene sets whose gene set members show coherent expression patterns. On the other hand, they propose to choose $p>1$ if the gene set database consists of large gene sets and only a small number of genes within the gene sets are expected to show a coherent expression behavior. In general, however, they recommend $p=1$ as a reasonable choice.

\subsubsection{PADOG}
In contrast to the remaining GSA methods, \tool{PADOG} does not contain any parameters that offer true flexibility to the user. While the user can specify a seed as well as the number of permutations in the assessment of the significance of a given gene set, it has already been discussed in Section `Indications and Recommendations for Users' from the main document that these parameters do not offer any true flexibility.

\section{Data Preprocessing Uncertainty}

\subsection{Gene ID Conversion}

\subsection{Removal of Duplicated Gene IDs} \label{duplicateID_removal}
We briefly review three approaches to removing duplicated gene IDs from a gene expression data set that are caused by the conversion of gene IDs. One way, which is the simplest approach considered here, is to remove the row of gene expression data among the duplicates that occurs first. This approach is used by \citet{silva2016tcga} and can be applied to both ways in which duplicates occur. \\
Secondly, to summarize all available information provided by the duplicated gene IDs, one can keep the (rounded) mean of all rows in the gene expression data set that correspond to the duplicated gene ID. However, this is only meaningful in the case when several different gene IDs from format A are converted to the same gene ID in format B (see Figure \ref{ID_duplicates} case 1) since then, there are several rows with distinct gene expression measurements mapped to the same ID in format B. In the second case, on the other hand, all rows from the gene expression data set with different gene IDs from format B, all of which are a result of a single gene ID in format A, contain identical gene expression measurements. This handling therefore corresponds to keeping one of the corresponding rows arbitrarily. \\
A third approach to removing duplicated gene IDs is to keep the row of the gene expression data set that contains the highest magnitude of count data across all samples. The intuition behind this is that a higher magnitude of counts is expected to result in a higher probability of the detection of differential expression and differential enrichment. This approach, again, only has a meaningful interpretation for the case in which several distinct gene IDs from format A are converted to format B.

\subsection{Pre-Filtering}
Different approaches to pre-filtering are particularly proposed alongside methods for differential expression analysis such as voom/limma, DESeq2, and edgeR. The approaches differ in complexity and in the magnitude of the number of genes that are excluded for further analysis. \\
Some methods, such as the methods \texttt{DESeq2} for differential expression analysis, suggest a simple and manual manner of pre-filtering in which those genes are excluded from further analysis that have less than a pre-specified number of read counts across all samples. Alternatively, a slightly more stringent approach is proposed in which only those genes are kept for subsequent analysis which have a pre-specified number of read counts in a pre-specified number of samples. These approaches particularly offer a certain amount of flexibility to the user since the threshold values can be chosen arbitrarily, however, it is reasonable to make the choice in accordance with the number of samples in the gene expression data set such that data sets with higher sample sizes are pre-filtered using a higher threshold value of read counts. \\
While this manner of pre-filtering disregards the effect of the library sizes of the different samples on the respective read counts, more sophisticated approaches to pre-filtering take this aspect into account. For instance, \texttt{edgeR} provides a function for pre-filtering which operates on the counts-per-million instead of the raw counts such that those genes are removed from further analysis that contain less than a pre-specified counts-per-million in a pre-specified number of samples. \\
Note that, while the user manual of \tool{GSEA} offers possible reasons why pre-filtering might be beneficial, they do not suggest any specific approach to pre-filtering. 

%\begingroup
% manipulate horizontal spacing
%\setlength{\tabcolsep}{7pt} % Default value: 6pt
% manipulate vertical Spacing
%\renewcommand{\arraystretch}{1.5} % Default value: 1

%\begin{landscape}

%\begin{table*}
%\small

   % \centering

   % \resizebox{\textwidth}{!}{%
   \renewcommand{\arraystretch}{1} % Default value: 1
    \begin{longtable}{p{0.17\linewidth} |Y{0.1\linewidth}Y{0.15\linewidth}p{0.45\linewidth}} 

    \caption{Overview of flexible choices in preprocessing for GSA. For each preprocessing step, it is indicated whether the user can choose between different approaches/methods and if a default (or particularly common) approach is available. The asterisk `*' indicates that the step is not truly flexible despite being specified by the user.}
         \label{flexible_adaptions_preprocess} \\

    \toprule

     \addlinespace[0.2cm]
     Preprocessing Step & Flexibility? & Default \newline approach available? & Description \\
     \midrule
     \addlinespace[0.2cm]
     Pre-filtering & \checkmark & \newcrossmark & The different approaches to pre-filtering are particularly proposed alongside methods for differential expression analysis such as voom/limma, DESeq2, and edgeR. The approaches differ in complexity and in the magnitude of the number of genes that are excluded for further analysis. \\
     Gene ID conversion & \checkmark & \newcrossmark & There are different \texttt{R} packages developed for the conversion of gene IDs between a variety of common formats. The conversion patterns between the packages are not uniform, leading to different gene IDs and different genes for whom no ID in the desired format exists. \\
     Removal of duplicated gene IDs & \checkmark & \newcrossmark & There exist no commonly accepted guidelines on approaches to remove duplicated gene IDs that result from the conversion of the gene IDs. A variety of ad-hoc approaches are described in statistics/bioinformatics forums, mostly based on personal experiences rather than formal evidence. \\
     Normalization \newline and transformation \newline (for FCS \textrm{I}) & \checkmark & \newcrossmark & While normalization is widely discussed in differential expression analysis and gene set analysis and a variety of methods exist, guidance on transformation is much scarcer and there are only a few methods proposed in the literature. \\

     Differential expression analysis \newline
     (for ORA and FCS \textrm{II}) & \checkmark & \newcrossmark & While there are a variety of available approaches which differ in aspects such as distributional assumptions, general approach, and method for normalization, a small subset of this variety, including voom/limma, DESeq2, and edgeR, is considerably more popular than the remaining methods. \\
     Gene-level statistic \newline (for FCS \textrm{II}) & \checkmark & \newcrossmark & There are only a few guidelines on valid metrics to rank the genes. Typically, user guidelines provided for the specific GSA methods advise the user to rank the genes by a suitable metric of their choice, leaving a great amount of flexibility to the user. Independent publications offer a small number of metrics, however, modifications of such metrics, which are mostly based on personal experience, are proposed copiously in online communities. \\
     Seed for reproducibility (for FCS) & * & * & While the seed can be specified by the user, it should be set arbitrarily and before inspecting the results and thus does not represent true flexibility. In particular, the seed should never be set to induce preferable results in a cherry-picking manner as these results would not be replicable on an independent dataset. \\
     
\bottomrule
    \end{longtable}%}
%\end{table*}

%\end{landscape}

%\endgroup

\subsection{Normalization}
We note that normalization is typically not performed as an individual step, but incorporated in other practical steps of the GSA workflow. While for ORA and FCS~II, normalization is usually included in the preceding differential expression analysis, it is typically performed as part of the transformation of the RNA-Seq data for FCS~I.

\subsection{Transformation}\label{subsec_transformation}
A transformation of the RNA-Seq data is applied such that the characteristics of the transformed RNA-Seq data are aligned with the distributional assumptions of the microarray methods. Two methods to transform (and additionally normalize) an RNA-Seq gene expression data set are described in the following. According to \citet{geistlinger2021toward}, these methods enable the application of microarray methods for the transformed RNASeq data. It is noted, however, that there seems to be no general consensus with respect to a method of transformation. \\

\subsubsection{Transformation using voom} \label{voom_transformation}
A common approach to normalizing and transforming an RNA-Seq expression data set is the log-cpm transformation employed by voom. 
The raw counts $K_{ij}$, $i=1,...,N$, $j=1,...,p$, are transformed by calculating the log-counts per million (log-cpm) as follows: 
\begin{align}
\label{voom_transf}
    \Tilde{K}_{ij}=\log_{2} \left( \frac{K_{ij} +0.5 }{S_{j}\cdot s_{j} + 1}\cdot 10^{6}\right),
\end{align}
where $S_{j}\cdot s_{j}$ is the library size of sample $j$ multiplied by the normalization factor, resulting in the effective library size. The normalization factor $s_{j}$ is obtained by performing a normalization technique. The authors of voom \citep{law2014voom} suggest using the trimmed mean of M-values method (`TMM') \citep{robinson2010scaling}, however, there again seems to be no general consensus on the choice of the normalization technique. The log-cpm values are by definition normalized for differences in library sizes across samples. By additionally scaling the library size $S_{j}$ with the Trimmed Mean of M-values method, the transformed values are normalized for compositionality effects. In Equation (\ref{voom_transf}), the raw count values $K_{ij}$ are offset by $0.5$ to reduce the variability of the log-cpm values for low counts and to avoid taking the $\log_{2}$ of 0. Additionally, the effective library size is offset by 1 to ensure that the fraction in the equation is strictly between 0 and 1.\\
Instructions found in user manuals \citep{geistlinger2021toward} and practical papers \citep{shahjaman2020robust,zhang2019novel} indicate that such cpm-transformed RNA-Seq measurements are assumed to be approximately normal distributed and hence to have a more stable mean-variance relationship. The precision weights, which complete the official use of voom, are, however, neglected. When applying the log-cpm transformation introduced in voom, but without the precision weights, it is important to bear in mind that the authors of voom \citep{law2014voom} argue that this `simple' approach may behave well in a scenario with a sufficiently high magnitude of counts but that it disregards the dependence between the variance and the mean for smaller magnitudes. \\

\subsubsection{Transformation using DESeq2's varianceStabilizingTransformation} \label{vst_transformation}
\label{vst}
Another transformation method for RNA-Seq data is integrated in the \texttt{DESeq2} package and denoted by \textit{varianceStabilizingTransformation} \citep{love2014moderated}. In this approach, the raw counts $K_{ij}$ are normalized using the median-of-ratios method \citep{anders2010differential}. The mean-variance relationship is estimated and the variance stabilizing transformation is then performed in a way that for transformed values, the variance is approximately constant throughout the range of the mean. Therefore, this transformation results in approximately homoscedastic values $\Tilde{K}_{ij}$ which are additionally normalized for library size. Furthermore, for large counts, the transformed values are asymptotically equal to the $\log_{2}$-values of the normalized counts.

\subsection{Differential Expression Analysis} \label{uncertainty_deAnalysis}
Among the parametric methods, DESeq2 and edgeR model the RNA-Seq data using the negative binomial distribution on which the statistical test is based. Both methods, however, differ in the technical aspects such as the method for normalization and the estimation of the dispersion. voom/limma, on the other hand, takes a different approach and incorporates characteristics specific to RNA-Seq data using voom into the differential expression pipeline limma that was initially developed for microarray data, making it applicable for RNA-Seq data.\\ In all of these parametric methods, differential expression is then assessed based on $p$-values, more specifically, adjusted $p$-values that account for the fact that a statistical test is performed for each gene in the experiment. \\

\subsubsection{DESeq2}
DESeq2 was developed specifically for RNA-Seq data and models the read counts using the negative binomial distribution. To account for different library sizes between samples and compositionality effects, normalization is performed using the median-of-ratios-method \citep{anders2010differential} which results in a separate normalization factor for each sample. To avoid the issue of unreliable log fold change estimates for genes with a low magnitude of counts, shrinkage of the log fold change values is performed. The resulting shrunken log fold change estimates can be directly used to rank genes in the reflection of their differential expression across the conditions.

\subsubsection{voom}
The voom approach is an extension to the standard limma pipeline and aims to utilize the normal-based statistical methods provided by standard limma for RNA-Seq data. In this context, special attention is given to the presence of the heteroscedastic nature of the count values in RNA-Seq data, i.e. the dependence of the variance in the count data on the mean of counts. The circumstance of heteroscedasticity is addressed by estimating the mean-variance relationship from the count data. This procedure results in a separate precision weight for each individual observation from the gene expression data. Together with the cpm-transformed count values, which are normalized for differences in library sizes across the samples, they are fed into the standard limma pipeline and an individual $p$-value of differential expression is computed for each gene from the experiment.

\subsubsection{edgeR}
Analogous to DESeq2, edgeR models the count data of each gene from the gene expression data set with the negative binomial distribution. To account for compositionality effects and differences in library sizes, an individual normalization factor for each sample is computed using the Trimmed Mean of M-values. The methodology behind edgeR places a focus on the estimation of the dispersion, especially caused by biological sources, to ensure an accurate assessment of differential expression of the genes. In the estimation of the dispersion for each individual gene, information from all genes in the experiment is utilized.

\subsection{Gene-Level Statistic (for FCS~II tools)}
The FCS~II tools \tool{GSEAPreranked} and GSEA conducted by \tool{clusterProfiler} require a list of all genes in the experiment as input which is ranked by the genes' corresponding values of the gene-level statistic. This value (or the rank of each gene within the list) then represents the magnitude of differential expression across the phenotypes and can be generated with a meaningful ranking metric of the user's choice. \citet{reimand2019pathway} suggest computing a gene ranking based on the gene-level statistic
\begin{align}
\label{ranking_pvalue}
    \text{gene-level statistic}= -\log_{10}(p) \cdot \text{sign}(\text{LFC}).
\end{align}
In this context, `$p$' denotes the unadjusted $p$-value and `LFC' refers to the log-transformed fold change between both conditions. Both quantities can be obtained from the results table of differential expression analysis performed with limma/voom, DESeq2, or edgeR. The intuition behind this ranking metric is that up-regulated genes, i.e. genes with a positive log fold change between both phenotypes, are assigned a positive rank, whereas down-regulated genes are allotted a negative rank. Moreover, a smaller $p$-value leads to a higher absolute rank compared to a large $p$-value. This means that, in sum, those genes with a higher absolute and significant fold change between the conditions are placed at the top or bottom of the ranked list.\\
In addition to this ranking metric that has been suggested by published literature, variants of it seem to be commonly used based on personal experience. \\
In contrast, users are suggested to use their `favorite t-test-like statistic' \citep{subramanian2005gene} in the user manual provided alongside the web-based tool \tool{GSEA}. For instance, an equivalent is provided in the results table generated with limma, namely the moderated t-statistic. \\
On the other hand, the authors of DESeq2 explicitly state that the shrunken log fold change estimates are suitable for generating a ranking of the genes for follow-up experiments \citep{love2014moderated}.

\subsection{Adjustment for Multiple Testing}
The intuition behind the necessity for multiple testing is that each individual test has a certain possibility of detecting false positive enrichment. The possibility of a false positive result then increases with the number of tests performed. To counteract this issue, the $p$-value of each test is systematically increased. This so-called `multiple testing adjustment' results in an adjusted $p$-value for each statistical test which is then used to assess differential enrichment of a given gene set. \\
One of the most common methods for adjusting for multiple testing is the procedure by Benjamini and Hochberg \citep{benjamini1995controlling}, which aims to control the expected proportion of false positive results among the positive results, called false discovery rate. Methods such as Benjamini and Hochberg assume independence between the null hypotheses. Other methods, such as the Bonferroni correction, do not assume independence between the hypotheses but are commonly known for their low statistical power when the number of tested hypotheses is high and/or when the hypotheses are dependent \citep{chen2017general}. Since there is a certain amount of overlap between the gene sets in terms of genes, independence is typically not given in the context of GSA. This can lead to inaccurate estimates of the false discovery rate \citep{reimand2019pathway} when a method for multiple testing adjustment is used that assumes independence. While this argument is in favor of Bonferroni and against Benjamini and Hochberg, many methods that perform multiple testing adjustment internally use the latter method by default (such as clusterProfiler and DAVID) nevertheless. In general, there yet seems to be no common default procedure for the adjustment of multiple testing in the context of gene set analysis that fully incorporates the dependence between the hypotheses. \\

\subsection{Gene Set Database}
While many GSA tools offer the usage of common gene set databases, the user can oftentimes also upload user-defined gene sets. Here, we restrict ourselves to common gene set databases.
\subsubsection{Gene Ontology (GO)}
Gene Ontology, abbreviated with `GO', summarizes up-to-date scientific knowledge about the functions of gene products such as RNA molecules or proteins resulting from the gene's expression \citep{ashburner2000gene,gene2023gene}. The overall goal is to understand in a species-independent way how individual genes contribute to the biology of an organism at the molecular, cellular, and organismal level. Gene Ontology is organized in the form of a directed acyclic graph in which each node corresponds to a specific gene set, also called `term'. In this context, a directed edge between two terms reflects their hierarchical relationship. In this sense, the `child' term is a subset of the `parent term'. The child term is more `specialized' than the parent term, meaning that its gene set members are related by a more specific relationship. The parent term, on the other hand, is more general. A direct consequence of this hierarchical relationship is that a gene from the child term is automatically also a member of the parent term, leading to a certain overlap between the gene sets.\\
Terms are categorized into three different subontologies, namely Molecular Function, Cellular Component, and Biological Process. The subontology Molecular Function entails terms in the form of activities of a gene product performed at the molecular level. In particular, a term does not describe the context, time, or location of the activity but instead only focuses on the activity itself. The molecular functions reported in this ontology mostly refer to activities performed by an individual gene product, yet some activities are performed by molecular processes composed of multiple gene products. An example of a broad functional term is `catalytic activity' or `transporter activity', whereas more specialized terms are `adenylate cyclase activity' or `toll-like receptor binding'.\\
%%%%%%%%
The subontology Cellular Component, on the other hand, refers to locations in which a gene product is active relative to the cellular structures whereas Biological Process describes pathways and larger processes to which a gene product's activity contributes. 
%These might either be cellular compartments, such as mitochondrion, or stable macromolecular complexes they are part of, e.g. ribosomes. 
%In contrast to terms of molecular functions, classes of cellular components do not refer to processes but instead to a cellular anatomy. \\
%This might be DNA repair or signal transportation in the broader sense, whereas examples of more specialized terms are pyrimidine nucleobase biosynthetic processes or glucose transmembrane transport. Due to the hierarchical structure in which GO is built, a gene corresponding to a child node is automatically annotated to the parent node and therefore, each gene corresponds to a multitude of GO terms. 

\subsubsection{Kyoto Encyclopedia of Genes and Genomes (KEGG)}
Kyoto Encyclopedia of Genes and Genomes \citep{kanehisa2000kegg, kanehisa2023kegg} is a resource of 16 databases which was initiated in 1995. These databases are classified into the four categories systems information, genomic information, chemical information, and health information. The purpose of KEGG is to impart an understanding of functions and utilities of the biological system on a higher level. Of interest in this paper is the database KEGG Pathways which is a collection of manually drawn pathway maps that represent the current knowledge of molecular interaction, reaction, and relation networks for metabolism, genetic information processing, environmental information processing, cellular processes, organismal systems, human diseases, and drug development. Each pathway map is a molecular reaction/interaction network diagram which ensures that experimental evidence in specific organisms can be generalized to other organisms through genomic information. In the following, KEGG Pathways will solely be referred to by the term `KEGG'.

\section{R Illustrations}
\begin{figure}[h]
     \centering
     \begin{subfigure}{0.9\textwidth}
         \centering
         \includegraphics[width=\textwidth]{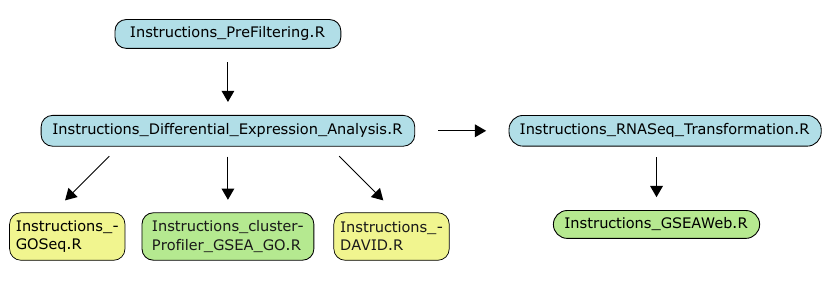}
         \caption{Sequence of \texttt{R} scripts for GSA tools that require a conversion of the gene IDs}
         \label{pipeline1}
     \end{subfigure}
     \hfill
     \begin{subfigure}{0.9\textwidth}
         \centering
         \includegraphics[width=\textwidth]{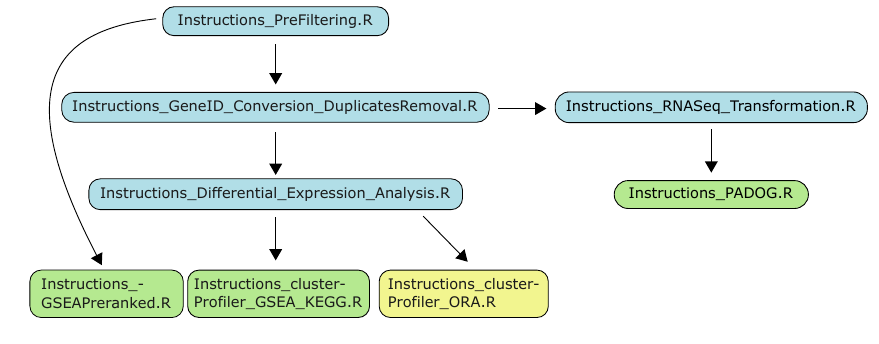}
         \caption{Sequence of \texttt{R} scripts for GSA tools that do not require a conversion of the gene IDs}
         \label{pipeline2}
     \end{subfigure}
     \caption{Overview of the sequence of \texttt{R} scripts that represent the individual steps from Figure 2 in the main document. }
     \label{fig_illustrations}
\end{figure}

A vital part of this paper is an illustrative \texttt{R} code to demonstrate the correct pipeline of the considered GSA tools. For the purpose of clarity, the \texttt{R} code is split up into several parts and the resulting structure is mainly aligned with Figure 2 from the main document. However, some discrepancies to this structure may occur if appropriate. For a visual illustration of the \texttt{R} codes, we provide a bookdown \citep{ref_bookdown} document in addition to the .R-files. This bookdown document follows the structure of the .R-files. The illustrations are based on the RNA-Seq measurements by \citet{pickrell2010understanding}, which are extracted from the lymphoblastoid cell lines of 69 unrelated Nigerian individuals \citep{pickrell2010understanding}. Since in this gene expression data set, the gene IDs are identified in the Ensembl format, a conversion of the gene IDs is required for those tools that do not accept this format. Refer to 
Table~\ref{accepted_geneIDs} to obtain an overview 
of the accepted gene ID formats per GSA tool. Correspondingly, Figure~\ref{fig_illustrations}, which provides an overview of the correct sequence of the \texttt{R} scripts for each GSA tool considered in this review, is split up into two parts. While Figure \ref{pipeline1} presents the correct sequence of \texttt{R} scripts for the set of GSA tools that accept the Ensembl ID format, \ref{pipeline2} provides the corresponding sequence for the tools which require a conversion of the gene IDs to an alternative format. Note that \tool{clusterProfiler}'s GSEA tool occurs in both Subfigures \ref{pipeline1} and \ref{pipeline2}. The reason for this is, as can be seen from Table \ref{accepted_geneIDs}, that when using gene set database KEGG, a conversion to Entrez gene ID is required, while for GO, Ensembl ID, as well as other formats, are accepted. \tool{GSEAPreranked} constitutes an exception to the overall structure of the \texttt{R}-files. Since it is the only tool that requires a conversion of the Ensembl gene IDs to HGNC gene symbols, a separate \texttt{R} script that follows pre-filtering ensures greater clarity.  \\

To run a sequence of \texttt{R} scripts, the working directory must be adapted to your computer using the command `setwd()'. In your working directory, the folder `data' is created to eventually contain the resulting objects of each corresponding script. In the left column of Table~\ref{results_folders}, an overview of the folders created in folder `data' throughout the \texttt{R} scripts is provided and supplemented by a short description of each file. \\
When running the sequence of \texttt{R} scripts for the first time, you have to run each script from the sequence separately and in the right order. Note that at the beginning of each script, the global environment is emptied to avoid confusion regarding objects with the same name but from different \texttt{R} scripts. Furthermore, when loading a library, i.e. \texttt{R} package, for the first time on your computer, you first have to install it using the code `install.packages()'. \\
At the end of each script,  there is a code line to save the resulting object in the corresponding file (see Table \ref{fig_illustrations}). After saving the resulting object, you can run the subsequent \texttt{R}-scripts independently since they contain a code line to reload it from the corresponding file. In combination with emptying the environment right at the beginning of each \texttt{R} script, this ensures that each script starts with only the relevant objects created in the preceding scripts. \\
At the end of each \texttt{R} script that illustrates the use of a GSA tool (see yellow and green nodes in Figure \ref{fig_illustrations}), an illustration is provided on how the parameters from Tables \ref{flexible_adaptions_ORA} and \ref{flexible_adaptions_FCS} can be adapted in the respective GSA tool.

\begin{table}[htbp]
\small
\centering
\caption{Description of folders that are created in folder `data'}

\label{results_folders}
\begin{tabular}{p{0.45\linewidth} |p{0.5\linewidth}}
\toprule
 Folder created in folder `data' &  Description \\
\midrule
Input\_Objects\_DAVID & To this file, the input objects for the web-based tool \tool{DAVID} are exported. They can be uploaded to the tool without further processing. \\
Input\_Objects\_GSEA\_web & To this file, the partly preprocessed input objects for the web-based tool \tool{GSEA}, containing the gene expression measurements and the conditions of the samples, are exported. Both require some further preprocessing steps in Excel.   \\
Input\_Objects\_GSEAPreranked & To this file, the partly preprocessed input object for the web-based tool \tool{GSEAPreranked}, namely the gene ranking, is exported. The ranking requires some further preprocessing in Excel.    \\ 
Results\_Differential\_Expression\_Analysis & To this file, the results of differential expression analysis that originate from DESeq2, limma/voom, and edgeR are exported. For each of the three methods, the results are exported for the genes in the Ensembl ID on the one hand and in Entrez ID on the other hand.  \\
Results\_GeneID\_Conversion\_DuplicatesRemoval & To this file, the (pre-filtered) gene expression data set with gene IDs converted to Entrez ID are exported. Since the methods for differential expression analysis differ in their proposed approach for pre-filtering, two differently pre-filtered gene expression data sets with genes in the Entrez format are exported.\\
Results\_PreFiltering & To this file, the pre-filtered gene expression data set with the gene IDs in the initial Ensembl ID format is exported. Note that two differently pre-filtered gene expression data sets are exported since the proposed approaches for pre-filtering differ between the methods for differential expression analysis. \\
Results\_RNASeq\_Transformation &  To this file, the transformed (and normalized as well as pre-filtered) gene expression data sets are exported. Since the tools \tool{GSEA} and \tool{PADOG} differ in the accepted gene ID format, a gene expression data set is exported in the Ensembl and Entrez ID, respectively. \\

\bottomrule
\end{tabular}
\end{table}